\documentclass{emulateapj}
\usepackage{natbib}
\usepackage{epsf}

\newenvironment{inlinefigure}{
\def\@captype{figure}
\noindent\begin{minipage}{0.999\linewidth}\begin{center}}
{\end{center}\end{minipage}\smallskip}

\slugcomment{ApJ, submitted}
\shorttitle{Leo IV SFH and Structure}
\shortauthors{Sand et al.}

\begin{document}
 \title{A Deeper Look at Leo IV: Star Formation History and Extended Structure$\!$\altaffilmark{1}}

\author{David J. Sand,$\!$\altaffilmark{2,3}  Anil Seth,$\!$\altaffilmark{3} Edward W. Olszewski,$\!$\altaffilmark{4} Beth Willman,$\!$\altaffilmark{5} Dennis Zaritsky,$\!$\altaffilmark{4} Nitya Kallivayalil$\!$\altaffilmark{6,7}} \email{dave.j.sand@gmail.com}

\begin{abstract}

We present MMT/Megacam imaging of the Leo~IV dwarf galaxy in order to
investigate its structure and star formation history, and to search
for signs of association with the recently discovered Leo~V satellite.
Based on parameterized fits, we find that Leo~IV is round, with
$\epsilon < 0.23$ (at the 68\% confidence limit) and a half-light
radius of $r_{h} \simeq 130$ pc.  Additionally, we perform a thorough
search for extended structures in the plane of the sky and along the
line of sight. We derive our surface brightness detection limit by
implanting fake structures into our catalog with stellar populations
identical to that of Leo~IV.  We show that we are sensitive to
stream-like structures with surface brightness $\mu_{r}\lesssim29.6$
mag arcsec$^{-2}$, and at this limit, we find no stellar bridge
between Leo IV (out to a radius of $\sim$0.5 kpc) and the recently
discovered, nearby satellite Leo V.  Using the color magnitude fitting
package StarFISH, we determine that Leo~IV is consistent with a single
age ($\sim$14 Gyr), single metallicity ($[Fe/H]\sim-2.3$) stellar
population, although we can not rule out a significant spread in these
value.  We derive a luminosity of $M_{V}=-5.5\pm0.3$.  Studying both
the spatial distribution and frequency of Leo~IV's 'blue plume' stars
reveals evidence for a young ($\sim$2 Gyr) stellar population which
makes up $\sim$2\% of its stellar mass.  This sprinkling of star
formation, only detectable in this deep study, highlights the need for
further imaging of the new Milky Way satellites along with theoretical
work on the expected, detailed properties of these possible
'reionization fossils'.

\end{abstract}
\keywords{} 

\altaffiltext{1}{Observations reported here were obtained at the MMT observatory, a joint facility of the Smithsonian Institution and the University of Arizona.}
\altaffiltext{2}{Harvard Center for Astrophysics and Las Cumbres Observatory Global Telescope Network Fellow}
\altaffiltext{3}{Harvard-Smithsonian Center for Astrophysics, 60 Garden Street, Cambridge MA 02138}
\altaffiltext{4}{Steward Observatory, University of Arizona, Tucson, AZ 85721}
\altaffiltext{5}{Haverford College, Department of Astronomy, 370 Lancaster Avenue, Haverford PA 19041}
\altaffiltext{6}{MIT Kavli Inst. for Astrophysics \& Space Research, 70 Vassar
 Street, Cambridge, MA 02139}
\altaffiltext{7}{Pappalardo Fellow}

\section{Introduction}

Since 2005, 14 satellite companions to the Milky Way have been
discovered (see \citet{willman09} and references therein).  Despite
the fact that many of these objects are less luminous than a typical
globular cluster ($-1.5 < M_V < -8.6$), these 14 objects have a range
of properties that encompass the most extreme of any galaxies,
including: the highest inferred dark matter content
\citep[e.g.][]{simongeha,geha09}, the lowest [Fe/H] content
\citep{Kirby08}, unusually elliptical morphologies
\citep[e.g. Hercules; ][]{Coleman07,sandherc}, and in some cases
evidence for severe tidal disturbance \citep[e.g. Ursa Major II;
][]{Zucker06,munoz09}.  The varied properties of these lowest
luminosity galaxies are valuable probes for understanding the physics
of dark matter and galaxy formation on the smallest scales.


Of the newly discovered Milky Way (MW) satellites, Leo~IV ($M_V =
-5.5$, $r_h \sim 130$ pc) is among the least studied, despite several
signs that it is an intriguing object.  Leo IV appears to be dominated
by an old and metal-pool stellar population \citep{sdsssfh}.  However
it also has an apparently complex color magnitude diagram (CMD), with
a 'thick' red giant branch, possibly caused by either multiple stellar
populations and/or depth along the line of sight \citep{Belokurov07}.
Leo IV also may have a very extended stellar distribution, despite its
apparently round ($\epsilon=0.22^{+0.18}_{-0.22}$) and compact
\citep[$r_{h}=2.5^{+0.5}_{-0.7}$ arcmin;][]{sdssstruct} morphology.  A
search for variable stars was recently performed by \citet{Moretti09},
who used the average magnitude of three RR Lyrae stars to find a
distance modulus of $(m-M)=20.94\pm0.07$ mag, corresponding to
$154\pm5$ kpc.  Interestingly, one of the three RR~Lyrae variables
lies at a projected radius of $\sim$10 arcmin, roughly three times the
half light radius, leading to the suggestion that Leo~IV may actually
possess a 'deformed morphology'.  

Based on Keck/DEIMOS spectroscopy of 18 member stars, Leo~IV has one
of the smallest velocity dispersions of any
of the new MW satellites, with $\sigma=3.3\pm1.7$ km/s
\citep{simongeha}.  A metallicity study of 12 of these spectra
\citep{Kirby08} showed Leo~IV to be extremely metal poor, with
$\langle [Fe/H] \rangle=-2.58$ with an intrinsic scatter of
$\sigma_{[Fe/H]}=0.75$ -- the highest dispersion among the new dwarfs.

Recently, the MW satellite Leo~V ($M_V \sim -4.3 \pm 0.3$) has been
discovered, separated by only $\sim$2.8 degrees on the sky and
$\sim$40 km/s from Leo~IV \citep{leov}. With a Leo~V distance of
$\sim$180 kpc, this close separation in phase space led \citet{leov}
to suggest that the Leo~IV/Leo~V system may be physically associated.
This argument was bolstered by \citet{leovspec}, who spectroscopically
identified two possible Leo~V members 13 arcminutes from the
satellite's center (Leo~V's $r_{h}$ is $\sim$0.8 arcminutes) along the
line connecting Leo~IV and Leo~V, suggesting that Leo~V is losing
mass.  A recent analysis by \citet{leoivleov} of two 1 square degree
fields situated between Leo~IV and Leo~V shows tentative evidence for
a stellar 'bridge' between the two systems with a surface brightness
of $\sim$32 mag arcsec$^{-2}$.


Motivated by all the above, we obtained deep photometry of Leo~IV with
Megacam on the MMT.  In this paper, we use these data to present a
detailed analysis of both the structure and SFH of Leo~IV.  We also
search for any signs of disturbance in Leo~IV which may hint at a past
interaction with the recently discovered, nearby Leo~V.  In
\S~\ref{sec:observations} we describe the observations, data reduction
and photometry.  We also present our final catalog of Leo~IV stars. In
\S~\ref{sec:struct} we derive the basic structural properties of
Leo~IV, and search for signs of extended structure.  We quantitatively
assess the stellar population of Leo~IV in \S~\ref{sec:starform} using
both CMD-fitting software and an analysis of its blue plume
population.  We discuss and conclude in \S~\ref{sec:discuss}.

\section{Observations and Data Reduction}\label{sec:observations}

We observed Leo IV on April 21 2009 (UT) with Megacam \citep{Megacam}
on the MMT.  MMT/Megacam has 36 CCDs, each with $2048\times4608$
pixels at 0\farcs08/pixel (which were binned $2\times2$), for a total
field-of-view (FOV) of $\sim$24'$\times$24'. We obtained 5 250s
dithered exposures in $g$ and in $r$ in clear conditions with 0\farcs9
seeing.  We reduced the data using the Megacam pipeline developed at
the Harvard-Smithsonian Center for Astrophysics by M. Conroy, J. Roll
and B. McLeod, which is based in part on M. Ashby's Megacam Reduction
Guide\footnote{http://www.cfa.harvard.edu/$\sim$mashby/megacam/megacam$\_$frames.html}.
The pipeline includes standard image reduction tasks such as bias
subtraction, flatfielding and cosmic ray removal.  Precise astrometric
solutions for each science exposure were derived using the Sloan
Digital Sky Survey Data Release 6 \citep[SDSS-DR6][]{sdsscite} and the
constituent images were then resampled onto a common grid (using the
lanczos3 interpolation function) and combined with
SWarp\footnote{http://astromatic.iap.fr/software/swarp/} using the
weighted average.

Stellar photometry was determined on the final image stack using a
nearly identical methodology as \citet{sandherc} with the command line
version of the {\sc DAOPHOTII/Allstar} package \citep{Stetson94}.  We
allowed for a quadratically varying PSF across the field when
determining our model PSF and ran {\sc Allstar} in two passes -- once
on the final stacked image and then again on the image with the first
round's stars subtracted, in order to recover fainter sources.  We
culled our {\sc Allstar} catalogs of outliers in $\chi^{2}$ versus
magnitude, magnitude error versus magnitude and sharpness versus
magnitude space to remove objects that were not point sources.  In
general, these cuts varied as a function of magnitude.  We
positionally matched our $g$ and $r$ band source catalogs with a
maximum match radius of 0\farcs5, only keeping those point sources
detected in both bands in our final catalog.

Instrumental magnitudes are put onto a standard photometric system
using stars in common with SDSS-DR6.  We used all SDSS stars within
the field of view with $17.5 < r < 21.5$ and $ -0.25 < g - r < 1.5 $
to perform the photometric calibration and simultaneously fit for a
zeropoint and a linear color term in $g-r$.  The linear color term
slope was 0.11 in ($r - r_{MMT}$) versus $(g-r)$ and 0.086 in
($g-g_{MMT}$) versus $(g-r)$, consistent with that found in the MMT
study of Bootes II by \citet{Walsh08}.  Slight residual zeropoint
gradients across the field of view were fit to a quadratic function
and corrected for (see also Saha et al. 2010), resulting in a final
overall scatter about the best-fit zeropoint of $\delta g \sim$0.05
and $\delta r \sim$0.05 mag.

To calculate our photometric errors and completeness as a
function of magnitude and color, we performed a series of artificial
star tests using a technique nearly identical to that of
\citet{sandherc}.  Briefly, artificial stars were placed into our Leo
IV images on a regular grid with spacing between ten and twenty times
the image full width at half maximum with the {\sc DAOPHOT} routine
{\sc ADDSTAR}.  In all, ten iterations were performed on our Leo IV
field for a total of $\sim$300000 implanted artificial stars.  The $r$
magnitude for a given artificial star was drawn randomly from 18 to 29
mag, with an exponentially increasing probability toward fainter
magnitudes.  The $g-r$ color is then randomly assigned over the range
-0.5 to 1.5 with equal probability.  These artificial star frames were
then run through the same photometry pipeline as the unaltered science
frames, applying the same $\chi^{2}$, sharpness and magnitude-error
cuts.  For reference, we are 50\% (90\%) complete in $g$ at $\sim$25.3
(23.6) mag and in $r$ at 24.8 (23.3) mag.  When necessary, such as for
calculating the SFH of Leo IV in \S~\ref{sec:starform}, the
completeness as a function of both magnitude and color is taken into
account.


\subsection{The Color Magnitude Diagram and Final Leo IV Catalog}\label{sec:fincat}

\begin{figure}
\begin{center}

\mbox{\epsfysize=10.0cm \epsfbox{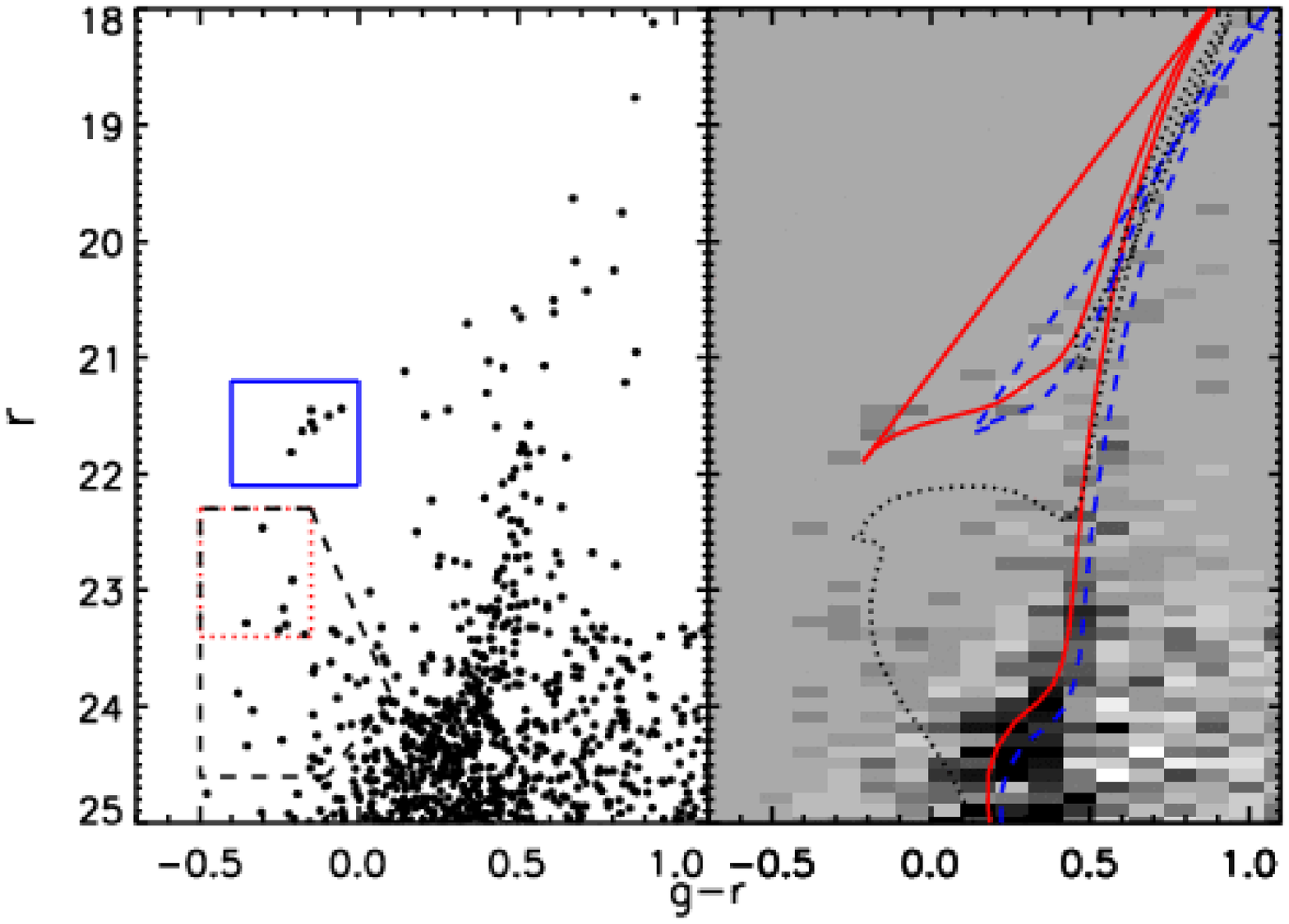}}

\caption{Left -- The CMD of Leo~IV within $r_{h}$=2.85 arcmin.  We
note several regions marking off particular stellar populations.  The
dash-lined region encloses the total blue plume population, although
any given star may be a foreground contaminant or unresolved galaxy.
The inset dotted box is our high probability blue plume box, with
relatively little galactic contamination.  Finally, we consider those
stellar objects within the solid-lined region to be BHB stars.  We
plot the spatial distribution of the high probability blue plume stars
and the BHBs in Figure~\ref{fig:showstars}.  Right -- A background
subtracted Hess diagram of the same, central half light radius region
of Leo~IV.  Overplotted are several theoretical isochrones from
\citet{Girardi04}.  The solid and dashed ridgelines are of a 14 Gyr
stellar population with [Fe/H]=-2.3, -1.7, respectively.  The dotted
line is a $\sim$1.6 Gyr isochrone with [Fe/H]=-1.3, and roughly
matches the overdensity of blue plume stars that are evident.
\label{fig:CMD}}
\end{center}
\end{figure}

We present the CMD of Leo~IV in Figure~\ref{fig:CMD}.  Plotted in the
left panel are all stars within the half-light radius (as determined
in \S~\ref{sec:structparams}), while in the right panel we present a
Hess diagram of the same field with a scaled background subtracted
using stars located outside a radius of 12 arcminutes.

In both panels we highlight the possible stellar populations of
Leo~IV.  In the right panel we plot three theoretical isochrones from
\citet{Girardi04}.  The solid and dashed lines are old 14 Gyr
isochrones with [Fe/H] of -2.3 and -1.7, respectively.  We adjust
these isochrones to have $m-M$=20.94, as found in the RR~Lyrae study
of Leo~IV by \citet{Moretti09}, which fit the ridgeline well.  With
the dotted line, we also plot a 1.6 Gyr isochrone with [Fe/H]=-1.3.
This isochrone agrees relatively well with the 'blue plume' stars
which are evident in Leo~IV's CMD.

We mark several regions in the left panel of Figure~\ref{fig:CMD} for
quantitative study in later sections.  The solid box denotes the blue
horizontal branch (BHB) stars, which are clearly defined in our CMD.
The dashed and dotted regions are two different selection regions for blue
plume stars.  The dashed is the total blue plume population, although
much of this region may be plagued by foreground stars or unresolved
galaxies.  We thus will also utilize the stars within the dotted box as a
relatively contamination-free tracer of the blue plume star population
in later sections.  An open question is whether or not the blue plume
stars are young stars, as is plausible based on the CMD, or are blue
stragglers.  We quantitatively assess the blue plume population in
\S~\ref{sec:BP}.

Given that the CMD of Leo~IV is visually in excellent agreement with
the distance measurement of \citet{Moretti09}, and the possible
presence of multiple stellar populations, we do not attempt a
CMD-fitting method for measuring the distance to Leo~IV
\citep[e.g.,][]{sandherc}, and will adopt $(m-M)=20.94$ throughout this
work.  We will vary this quantity when necessary to determine how
sensitive our results are to this assumption.

We present our full Leo~IV catalog in Table~\ref{table:phot}, which
includes our $g$ and $r$ band magnitudes (uncorrected for extinction)
with their uncertainty, along with the Galactic extinction values
derived for each star \citep{Schlegel98}.  We also note whether the
star was taken from the SDSS catalog rather than our MMT data, as was
done for objects near the saturation limit of our Megacam data.
Unless stated otherwise, all magnitudes reported in the remainder of
this paper will be extinction corrected.

\section{Leo IV Structural Properties}\label{sec:struct}

We split our analysis of the structural properties of Leo~IV into two
components.  First, we fit parameterized models to the surface density
profile of Leo~IV.  Following this, we search for signs of extended
structure in Leo~IV, especially in light of its proximity to Leo~V.

\subsection{Parameterized Fit}\label{sec:structparams}

\begin{figure}
\begin{center}
\mbox{\epsfysize=10.0cm \epsfbox{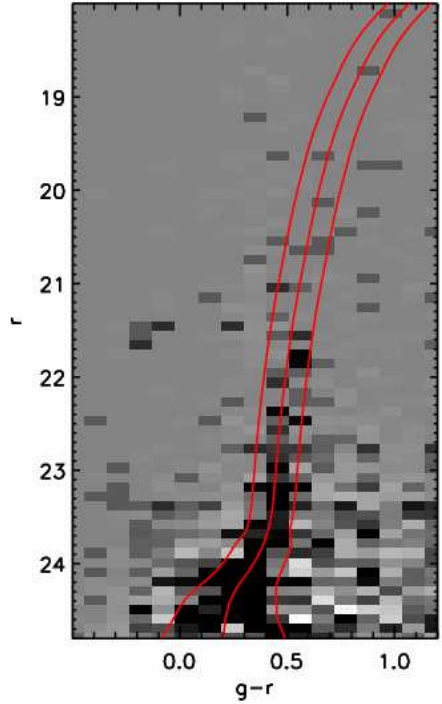}}
\caption{An illustration of the CMD selection region used for
performing our parameterized structural analysis of Leo~IV.  The
figure is a background-subtracted Hess diagram of the central 2.5
arcminutes of Leo~IV and the central, red fiducial ridge line is that
of M92 transformed to a distance modulus of $(m-M)=20.94$.  The two
bordering ridge lines bound the selection region and are a minimum of
0.1 magnitudes in the color direction away from the M92 ridge line.
The selection region expands to the size of the typical $g-r$ color
uncertainty as a function of $r$ mag when this exceeds 0.1 mag, as
determined via our artificial star tests.  The selected region is an
excellent match to the apparent ridgeline of Leo~IV, as can be seen in
the figure.
\label{fig:CMDselect}}
\end{center}
\end{figure}

It is common to fit the surface density profile of both globular
clusters and dSphs to King \citep{King66}, Plummer \citep{Plummer11},
and exponential profiles.  This is an important task both to
facilitate comparisons with other observational studies and for
understanding the MW satellites as a population
\citep[e.g.,][]{sdssstruct}.  To do this, we use the CMD selection
region shown in Figure~\ref{fig:CMDselect} for isolating likely Leo IV
members.  This CMD selection box was determined by first taking a M92
ridgeline at $(m-M)$=20.94, the distance to Leo~IV found by
\citet{Moretti09}, and placing two bordering selection boundaries a
minimum of 0.1 mag on either side in the $g-r$ color direction.  These
selection regions are increased to match the typical $g-r$ color
uncertainty at a given $r$ magnitude (as determined with our
artificial star tests) when that number exceeds 0.1 mag.  A magnitude
limit of $r$=24.8 mag was applied to correspond to our 50\%
completeness limit.  At the present we focus on Leo IV ridge line
stars, but we discuss the spatial properties of the blue plume and
horizontal branch stars in \S~\ref{sec:BP}.

We fit the three parameterized density profiles to the
stellar distribution of Leo IV:

\begin{equation}
\Sigma_{King}(r) = \Sigma_{0,K}\left( \left(1+\frac{r^2}{r_c^2}\right)^{-\frac{1}{2}}-\left(1+\frac{r_t^2}{r_c^2}\right)^{-\frac{1}{2}}\right)^2
\end{equation}

\begin{equation}
\Sigma_{Plummer}(r) =  \Sigma_{0,P}\left(1+\frac{r^2}{r_P^2}\right)^{-{2}}
\end{equation}

\begin{equation}
\Sigma_{exp}(r) =  \Sigma_{0,E}\exp\left(-\frac{r}{\alpha}\right)
\end{equation}

\noindent where $r_P$ and $\alpha$ are the scale lengths for the
Plummer and exponential profiles and $r_c$ and $r_t$ are the King core
and tidal radii, respectively. For the Plummer profile, $r_P$ equals
the half-light radius $r_h$, while for the exponential profile $r_h
\approx 1.668\alpha$.  We simultaneously fit a background surface
density, $\Sigma_{b}$, while fitting the Plummer and exponential
profiles.  For the King profile, there is a degeneracy between the
tidal radius and the background surface density.  We thus fix the
background value to the average of that found for the Plummer and
exponential profiles for our King profile fits
\citep[e.g.][]{Walsh08}.

We use a maximum likelihood (ML) technique for constraining structural
parameters similar to that of \citet{sdssstruct}, which we have
refined in \citet{sandherc}, and further refined in the current
work. We point the reader to those works for further details
concerning the expression of the likelihood function.  Including the
central position, $\alpha_{0}$ and $\delta_{0}$, position angle
($\theta$), and ellipticity ($\epsilon$) both the exponential and
Plummer profiles have the same free parameters --
($\alpha_{0}$,$\delta_{0}$,$\theta$,$\epsilon$,$r_{half}$,$\Sigma_{b}$),
while the King profile free parameters are
($\alpha_{0}$,$\delta_{0}$,$\theta$,$\epsilon$,$r_{c}$,$r_{t}$).
Uncertainties on structural parameters are determined through 1000
bootstrap resamples, from which a standard deviation is calculated.
We have tested the robustness of our algorithm for dwarf galaxies with
roughly the same number of stars as Leo IV in an Appendix, and will
use these tests to inform our results in what follows.

Our results are presented in Table~\ref{table:paramfits}.  We show our
best fit stellar profiles in Figure~\ref{fig:stellarprofile}.
Although the plotted stellar profiles are not fit to the plotted
binned data points, they do show excellent agreement.  We note that
the apparent slight overdensity at $R\sim$8 arcmin above the derived
parameterized fits does not correspond to any single feature, as can
be seen from the smoothed map of Leo~IV that we present in
\S~\ref{sec:extend} and Figure~\ref{fig:smoothmap}, but is likely just
the result of several fluctuations at roughly the same radius.
Interestingly, Leo~IV appears to be particularly round, at least
according to the parameterized model fit to the data.  In fact, our
ML-derived ellipticity is consistent with zero (see also the
Appendix), allowing us to only place an upper limit of $\epsilon
\lesssim 0.23$ (at the 68\% confidence limit).  Given this low
ellipticity, we can not place any meaningful constraints on the
position angle of Leo IV, as both the tests in the Appendix indicate
and the bootstrap resamples of our Leo IV data reaffirm.  On a
separate note, the tidal radius for the King profile fit, with a value
of $r_t=18.55'$, is larger than the Megacam FOV.  Thus, this value
should be taken with caution.

It is useful to compare our parameterized fit results with similar
work in the literature.  Using the original SDSS data,
\citet{sdssstruct} fit Leo~IV with an exponential profile using a ML
technique similar to the one utilized in the current work, and found
results within 1-$\sigma$ of those presented here.  More recently,
\citet{leoivleov} used deeper data from the 3.5 m Calar Alto telescope
around both Leo~IV and Leo~V, again applying a ML algorithm to measure
their structure.  In this case, the authors find $r_h=4.6^{+0.8}_{-0.7}$
and a well measured ellipticity of $\epsilon=0.49\pm0.11$, both
statistically inconsistent with the results presented in the current
work.  As commented on by \citet{leoivleov}, this is likely because of
the different stellar populations probed -- while the present work
uses mostly main sequence and subgiant stars to determine the
structure of Leo IV, \citet{leoivleov} use mostly brighter stars and
objects on the blue horizontal branch.  Also, Leo~IV lies on the edge
of one of their pointings, which could have biased their results.
Resolution of this discrepancy will require additional measurements.


\subsection{Extended Structure Search}\label{sec:extend}

We now search for signs of tidal disturbance and other anomalies --
which would not be picked up by our parameterized fits in
\S~\ref{sec:structparams} -- based on the morphology of Leo~IV's
isodensity contours.  We do this with an eye towards determining if
there is a current physical connection between Leo~IV and Leo~V.

\begin{figure*}
\begin{center}
\mbox{
\epsfysize=5.0cm \epsfbox{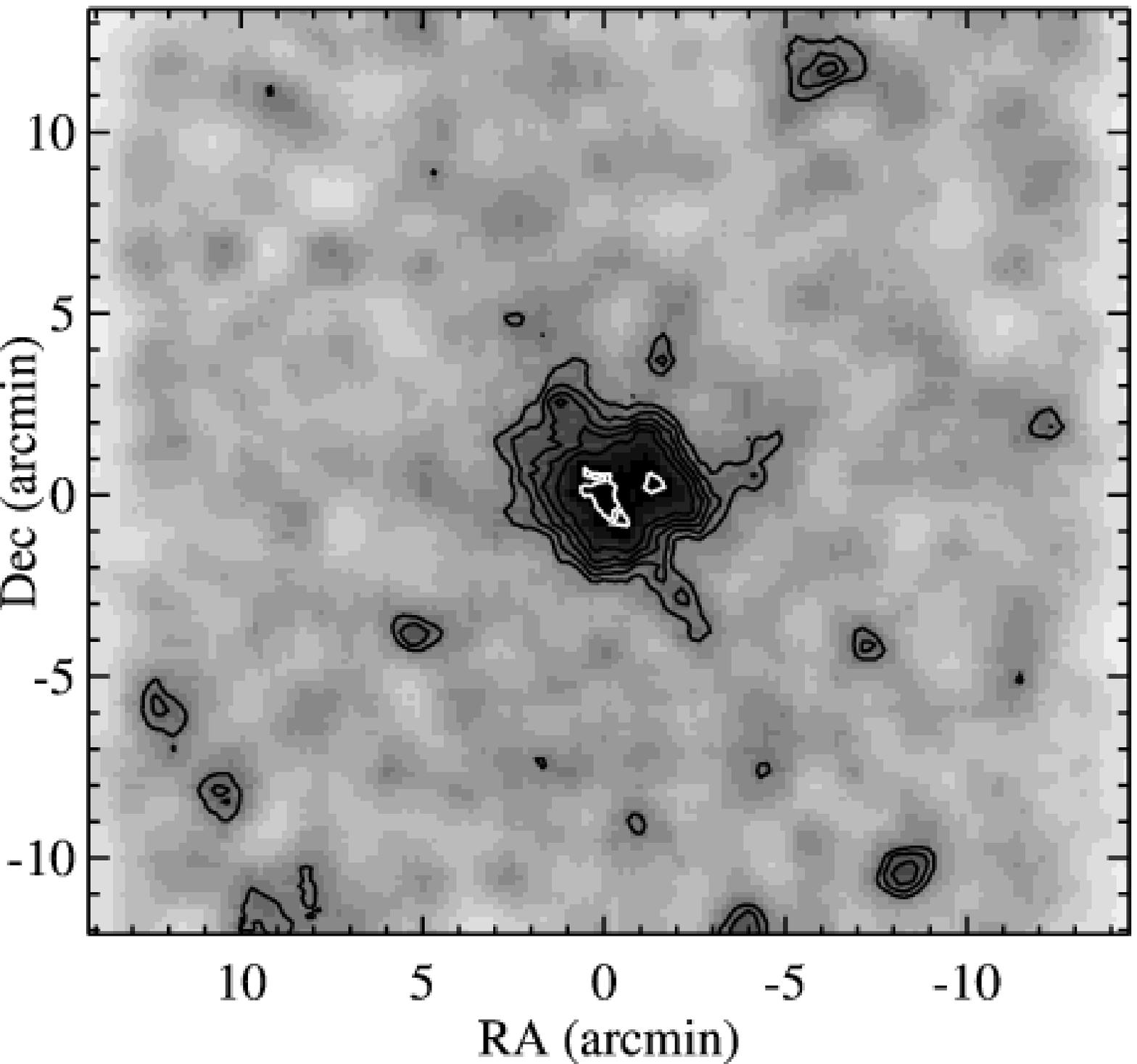}
\epsfysize=5.0cm \epsfbox{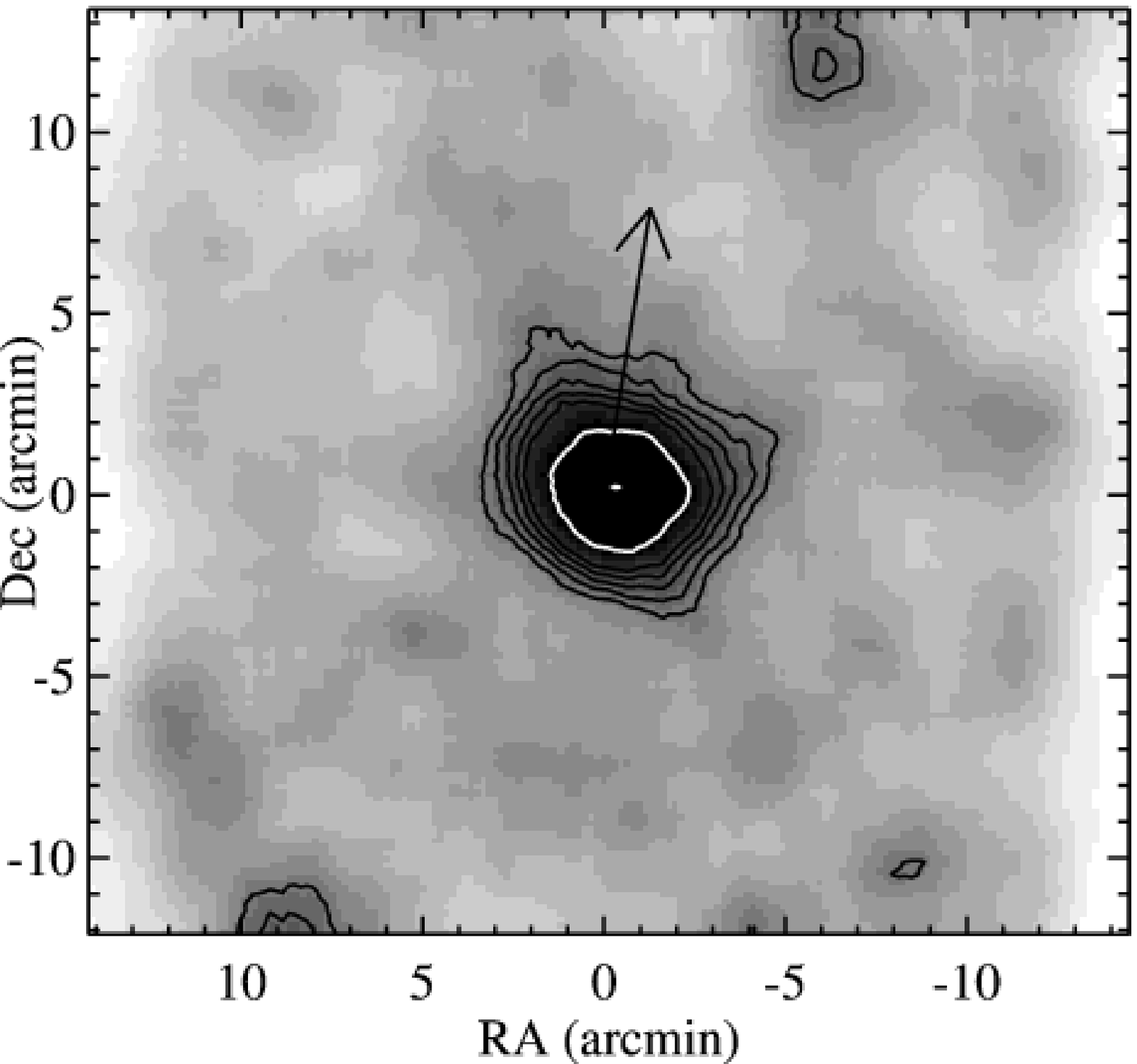}
\epsfysize=5.0cm \epsfbox{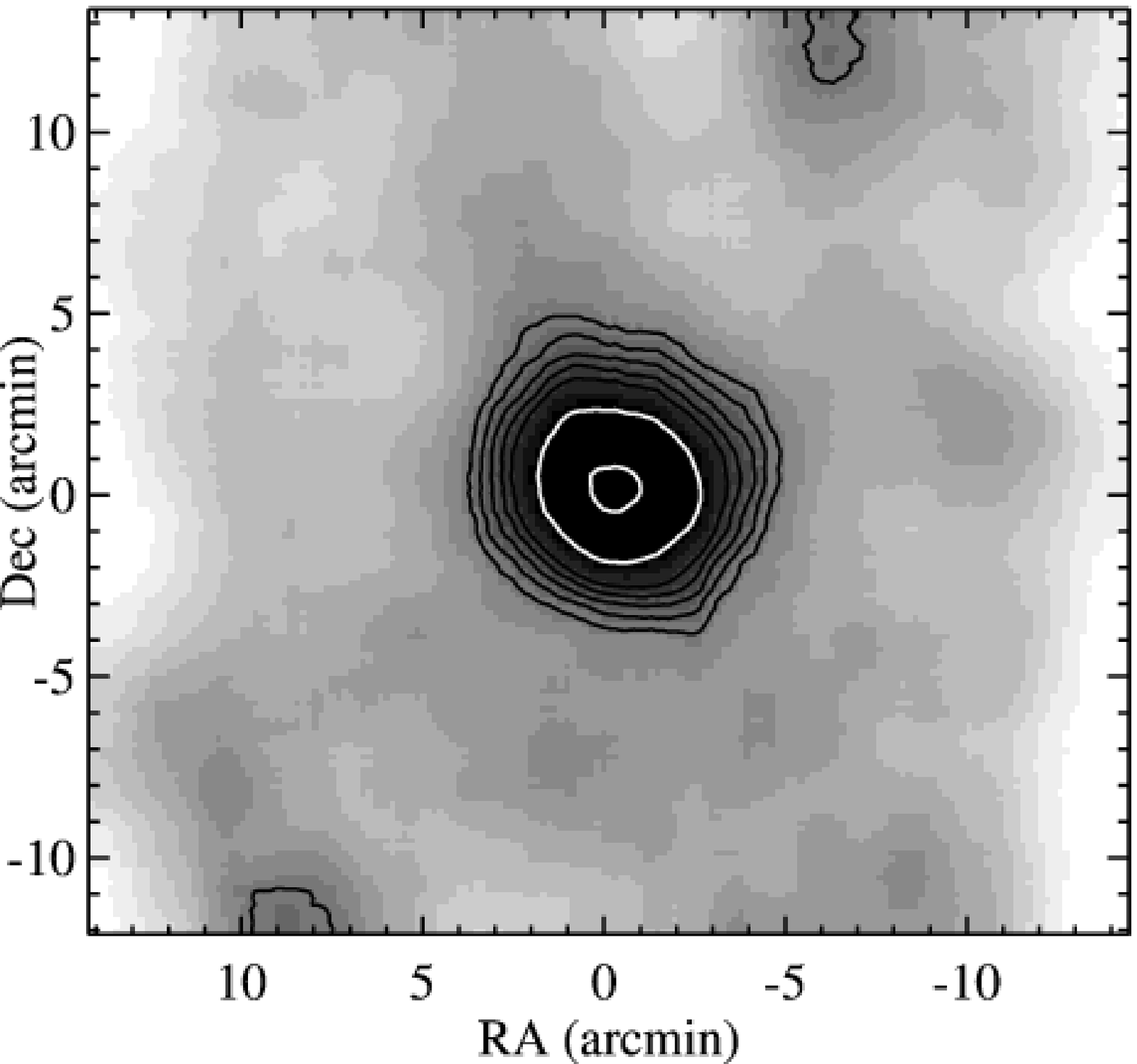}
}

\caption{ Smoothed contour plots of Leo~IV, showing the 3, 4, 5, 6, 7,
10 and 15 $\sigma$ levels.  The left panel has been smoothed by 0.5
arcmin, the middle panel by 1.0 arcmin, and the right panel by 1.5
arcmin.  The arrow in the middle panel indicates the direction to
Leo~V.  Note that there is no apparent structure along this direction.
The 3 $\sigma$ contour in our 0.5 arcminute smoothed map corresponds
roughly to a surface brightness of $\sim$30.0 mag per arcsecond$^{2}$.
\label{fig:smoothmap}}
\end{center}
\end{figure*}

Our basic approach is similar to that of \citet{sandherc}.  We include
stars within the same CMD selection box used for our parameterized
fits (Figure~\ref{fig:CMDselect}), placed those stars in
$10''\times10''$ bins and spatially smoothed these pixels with three
different Gaussians of $\sigma$=0.5,1.0 and 1.5 arcminutes.  The
background level of stars in the CMD selection box, and its variance,
was determined via the MMM routine in IDL\footnote{available at
http://idlastro.gsfc.nasa.gov/}.  To avoid the bulk of Leo~IV, these
statistics were determined in two boxes of size $20'\times3'$ situated
9.5 arcminutes North and South of Leo~IV.  We found that our smoothed
maps were unaffected if only one of these boxes was used, or if we
varied their sizes. We present our smoothed maps in
Figure~\ref{fig:smoothmap}, with the contours representing regions
that are 3, 4, 5, 6, 7, 10 and 15 standard deviations above the
background.  We focus on the $\sigma$=0.5 arcminute map, since it
contains the most detail without loss of potential Leo~IV features.
The $\sigma$=1.0 arcminute smoothing scale will be useful in
\S~\ref{sec:fake}, as we explore our sensitivity to stellar streams.

\begin{figure*}
\begin{center}
\mbox{
\epsfysize=15.0cm \epsfbox{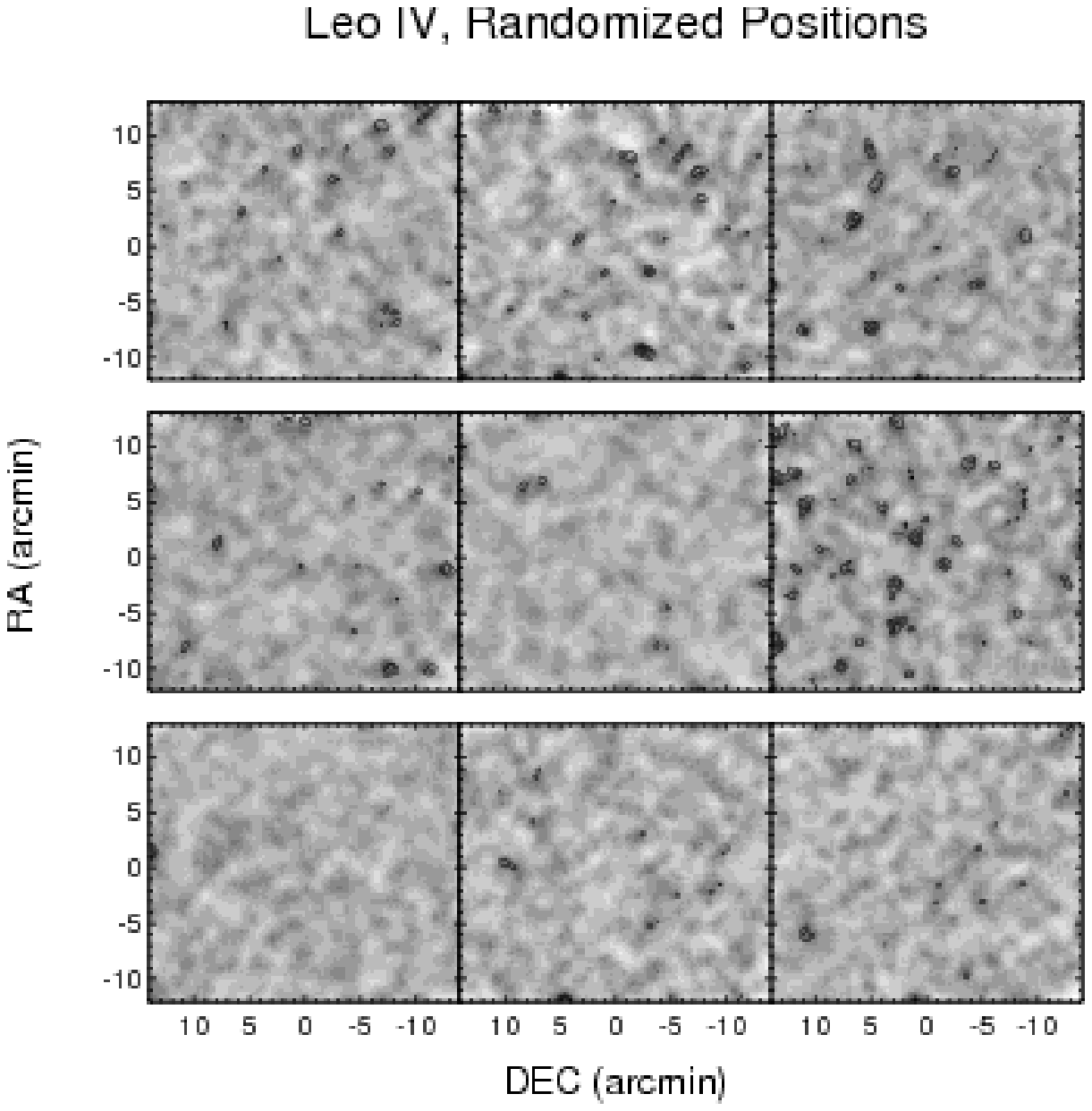}
}

\caption{Smoothed contour plots of nine random realizations of
Leo~IV stars, where we have reassigned star positions randomly across
the Megacam field of view.  The contours show the 3, 4 and 5
$\sigma$ levels.  The plots show that compact 3 and 4  $\sigma$
overdensities are relatively common.
\label{fig:randompos} }
\end{center}
\end{figure*}

Outside of the main body of Leo~IV, there appears to be only a handful
of compact, 3, 4 and 5 $\sigma$ overdensities.  How significant are
these overdensities, given the binning, smoothing and number of pixels
that went into the making of Figure~\ref{fig:smoothmap}?  To gauge
their significance, we take our input photometric catalog, randomize
the star positions and remake our smoothed maps (with the 0.5 arcmin
Gaussian) for several realizations (Figure~\ref{fig:randompos}).
While some of these maps have several 3 and 4 $\sigma$ overdensities,
while others have fewer, we find that the distribution of pixel values
maintains a Poisson distribution.  We thus conclude that the majority
of features outside the main body of Leo~IV in
Figure~\ref{fig:smoothmap} are likely just noise -- with the possible
exception of the 5-$\sigma$ overdensities at positions
$(\sim-6',\sim12')$ and $(\sim-8',\sim-10')$ with respect to Leo~IV.
Background-subtracted Hess diagrams of these two regions were made
from our Leo~IV catalog, but they do not yield CMDs that are
consistent with Leo~IV's stellar population (see \S~\ref{sec:fake}).
Thus, our observations yield no strong evidence for substructure in
the vicinity of Leo~IV.

\begin{figure*}
\begin{center}
\mbox{
\epsfysize=15.0cm \epsfbox{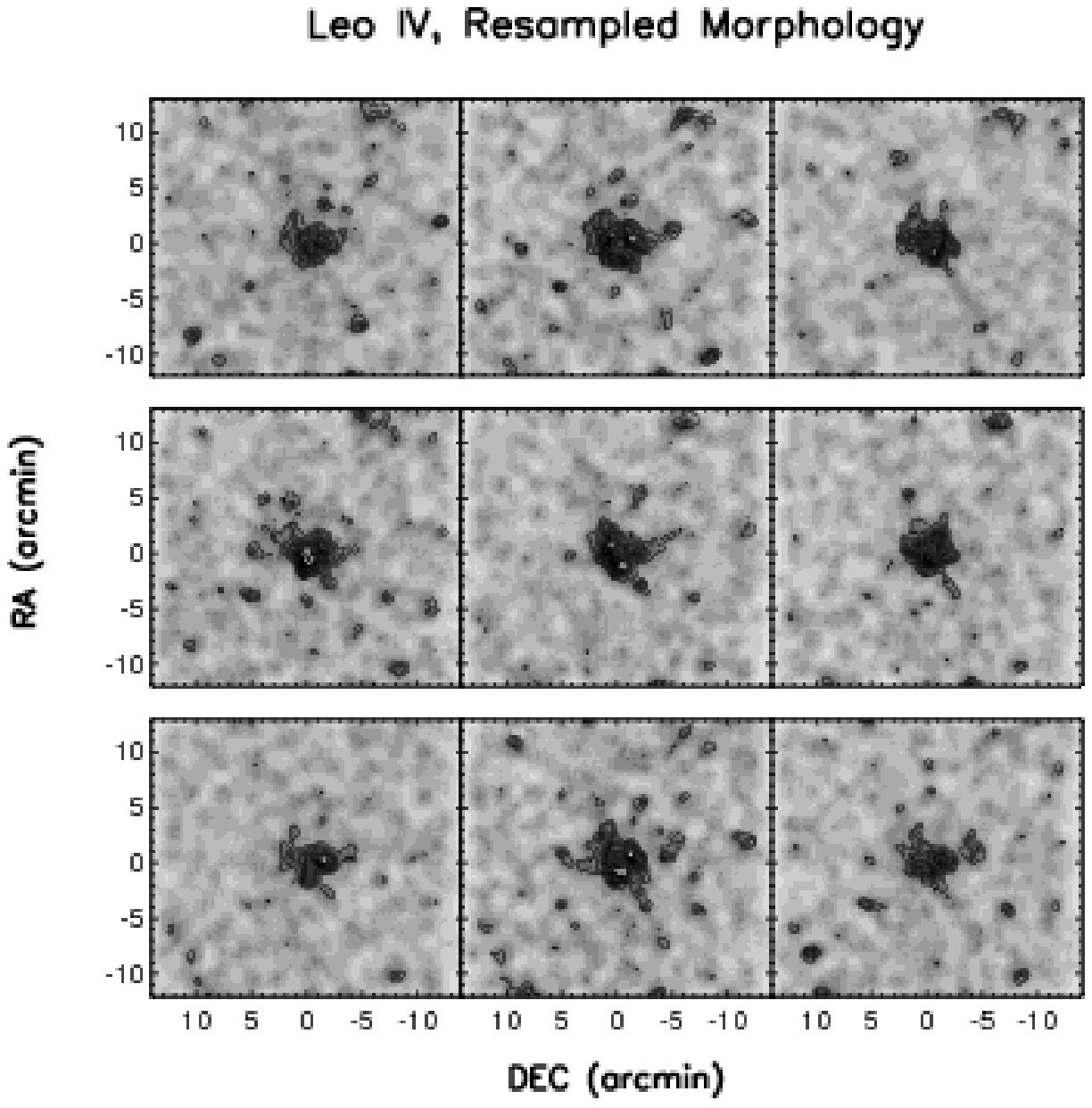}
}

\caption{ Smoothed contour plots of 9 random bootstrap resamples of
Leo~IV stars (with the 0.5 arcminute Gaussian), showing how morphology
can vary with each iteration.  The contours show the 3, 4, 5, 6, 7, 10
and 15 $\sigma$ levels.  While in some iterations the 'tendrils' seen
in Figure~\ref{fig:smoothmap} are still visible, they are not a
ubiquitous feature.  Future, deeper imaging is necessary to confirm
their reality.
\label{fig:resampmaps}}
\end{center}
\end{figure*}

The main body of Leo~IV itself has some interesting features.  There
is a hint of an elongation or disturbance in the core of Leo~IV, along
with two 'tendrils' -- one directed to the West and the other to the
Southwest.  Again, due to the small number of stars in Leo~IV, these
irregularities may be an effect of small number statistics.  To
evaluate the significance of the morphology in
Figure~\ref{fig:smoothmap}, we follow the path of \citet{Walsh08} and
their evaluation of morphological irregularities in Bootes~II.  We
bootstrap resample the Leo~IV stars and replot our smoothed maps.  The
results of nine such resamples can be seen in
Figure~\ref{fig:resampmaps}.  While we cannot rule out that the
tendrils seen in our Leo~IV map are genuine, they are not ubiquitous
features in our resampled maps, and so we cannot with confidence claim
they are real features.

We finally point out that there appears to be no sign of interaction
or disturbance in the direction of Leo~V, which we indicate in the
middle panel of Figure~\ref{fig:smoothmap}.  A Hess diagram of all
stars farther than 5 arcminutes from the center of Leo~IV, and within
1.5 arcminutes of the line that connects Leo~IV and Leo~V, yields a
differenced CMD that is consistent with noise (after proper background
subtraction).  Any such disturbance would have to be below our
detection threshold, which we determine in the next section.  Finally
we point out that these smoothed maps are sensitive to structures at
the distance of both Leo~IV and Leo~V ($\sim$180 kpc; Belokurov et
al. 2009) given that we are predominantly probing the CMD at
magnitudes brighter than the subgiant branch.  


\subsubsection{Inserting Artificial Remnants}\label{sec:fake}

In order to assess our surface brightness limit, and our sensitivity
to structures of different sizes and morphologies, we insert fake
'nuggets' and 'streams' into our Leo~IV catalog with stellar
populations drawn from one consistent with that of Leo~IV.  As in
\citet{sandherc}, we use the {\it testpop} program within the
CMD-fitting package, StarFISH, to generate our artificial 'Leo~IV'
CMDs and then remake our smoothed maps.  By varying the number of
stars (and, by extension, the surface brightness) in these structures,
we can then assess whether our adopted search method for extended
structure would have recovered them, and if so, what the resulting CMD
would look like.  This method of 'observing' and then examining these
artificial remnants is more informative than traditional methods of
simply quoting a '$3-\sigma$' surface brightness limit, even if the
result is a relatively ambiguous detection limit.  For details of the
procedure, we refer the reader to \citet{sandherc}, and to our
presentation of Leo~IV's SFH in \S~\ref{sec:starform}.

\begin{figure*}
\begin{center}
\mbox{
\epsfysize=8.0cm \epsfbox{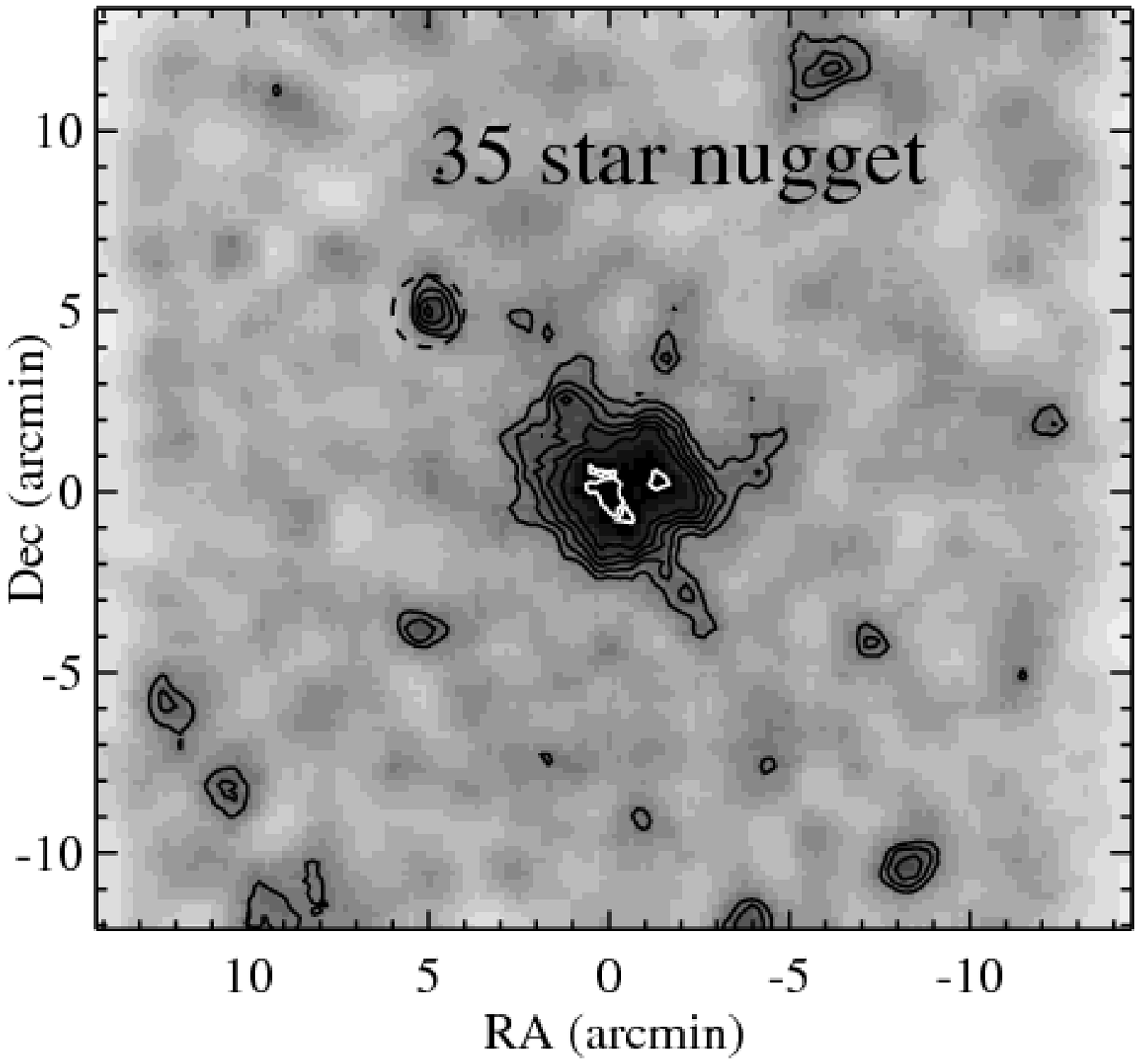}
\epsfysize=6.0cm \epsfbox{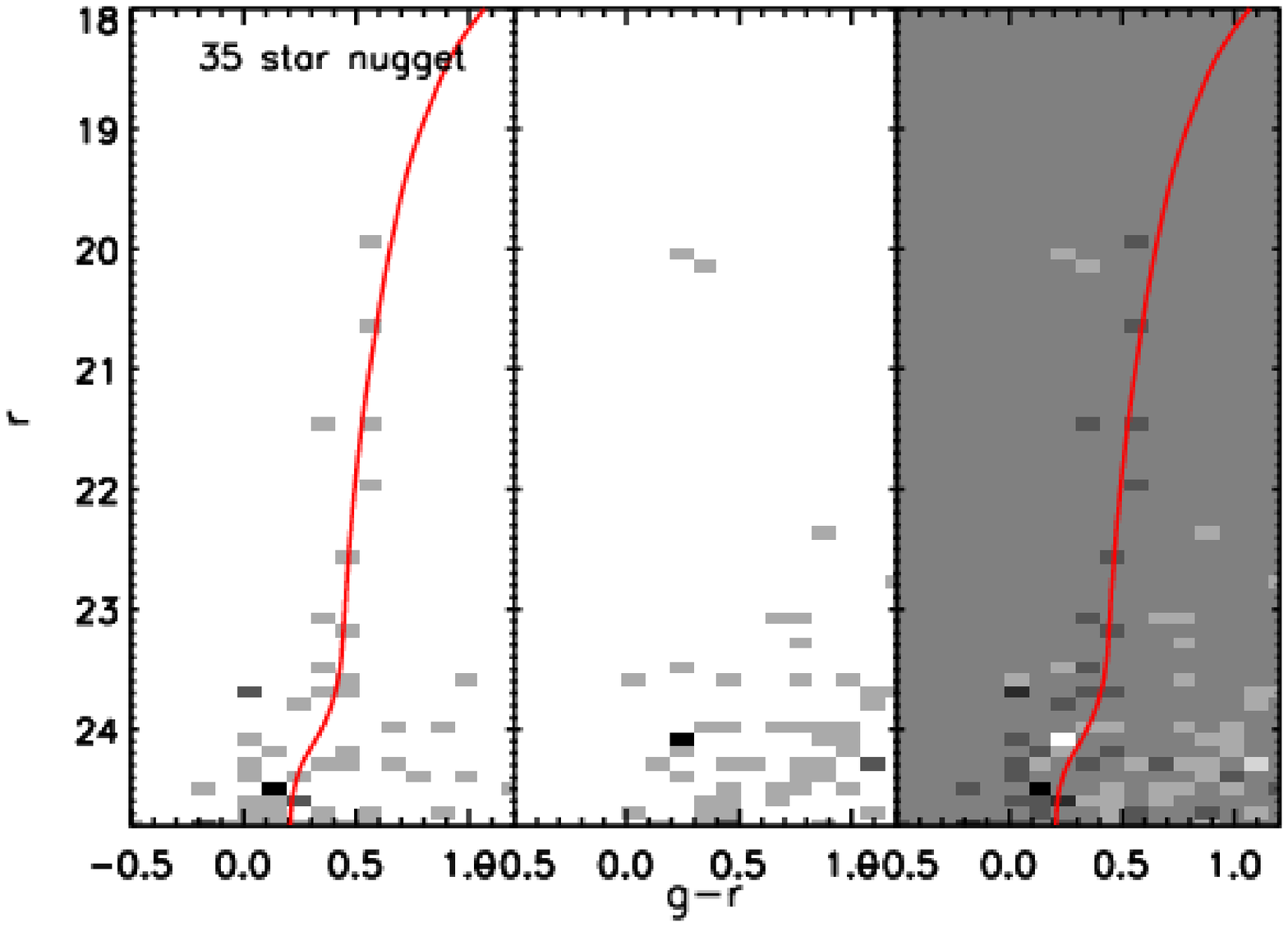}
}

\caption{An illustration of our technique for implanting fake Leo~IV
'nuggets' into our field, using an artificial stellar population drawn
from that determined in our star formation history analysis
(\S~\ref{sec:starfish}) and our artificial star tests
(\S~\ref{sec:observations}).  Left: We show our smoothed map of Leo~IV
after implanting a 35 star 'nugget' at $(+5',+5')$ with respect to the
center of Leo~IV, distributed as an exponential profile with a
half-light radius of 1 arcminute (see Table~\ref{tab:fakeresult}).
This nugget results in a $\sim$6.4-$\sigma$ overdensity at that
position.  Right: A Hess diagram of the 35 star nugget, along with a
M92 isochrone shifted to $(m-M)=20.94$.  This Hess diagram shows the
raw star counts in the circular aperture shown in the left panel of
this figure, along with an equal area background region, and the
difference between the two.  Note that the residual CMD shows several
stars that would be identified with Leo~IV's red giant branch, which
would allow us to say with confidence that such an overdensity is
likely associated with Leo~IV.
\label{fig:nuginj}}
\end{center}
\end{figure*}

\begin{figure*}
\begin{center}
\mbox{
\epsfysize=5.5cm \epsfbox{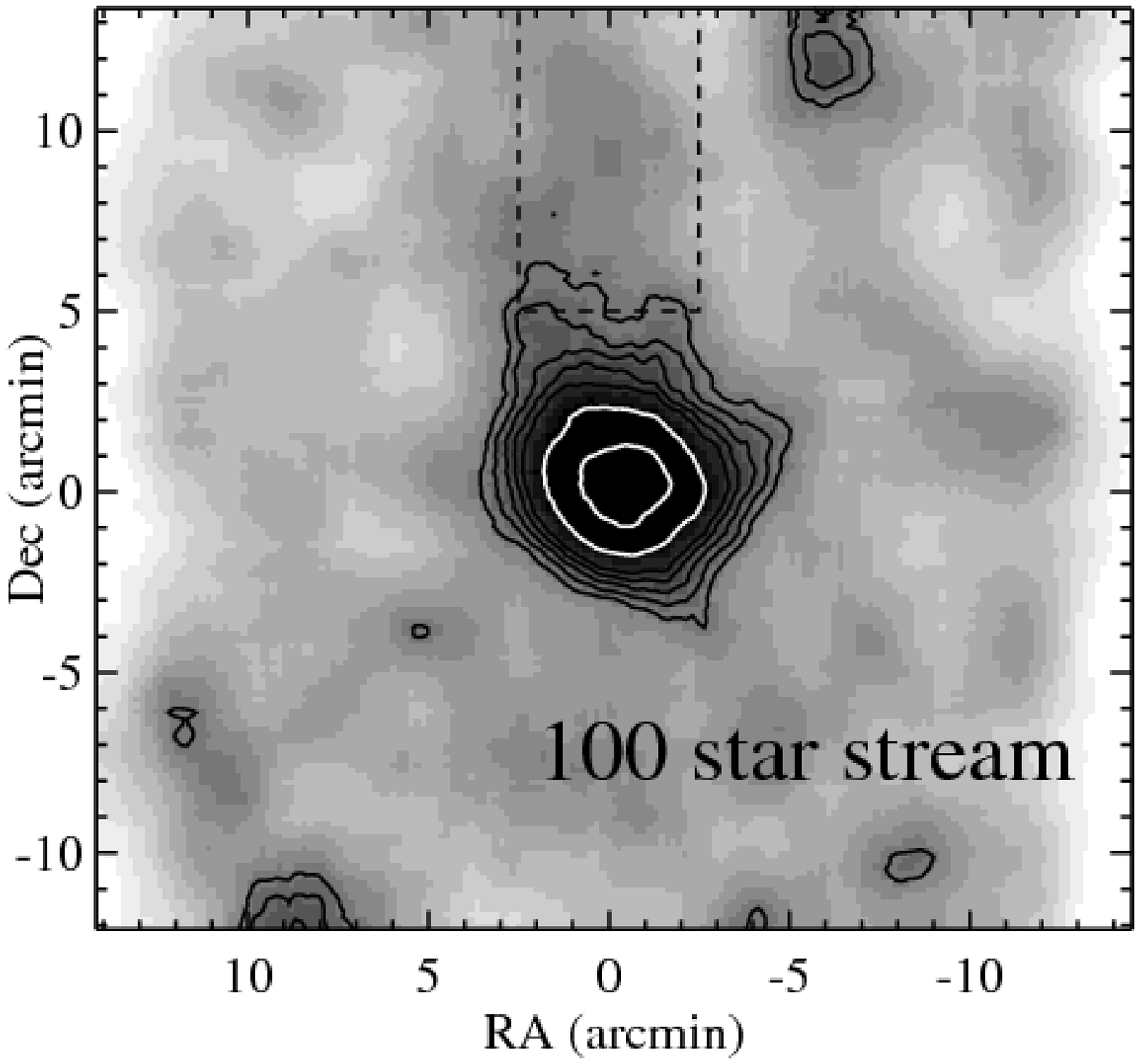}
\epsfysize=5.5cm \epsfbox{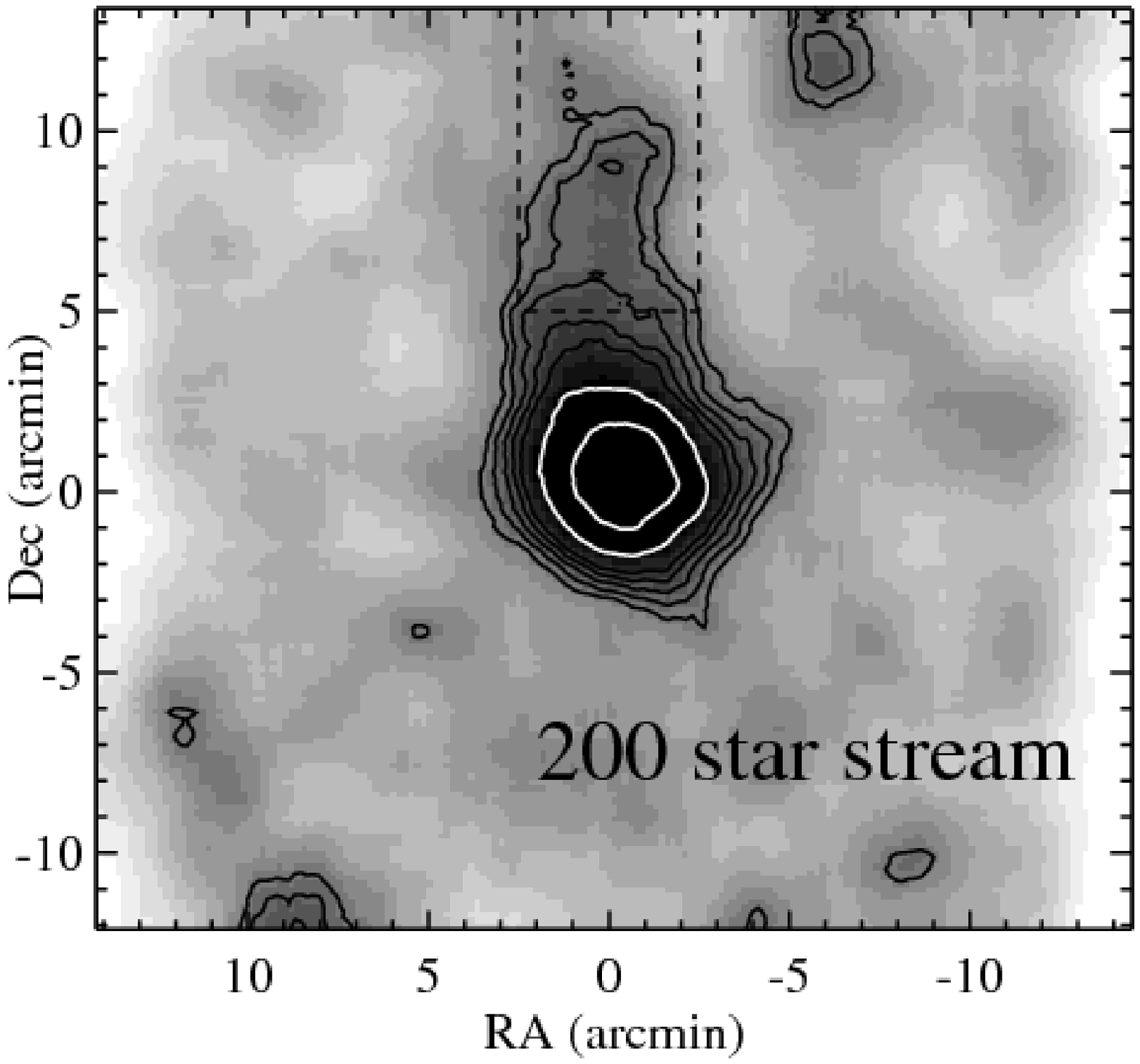}
\epsfysize=5.5cm \epsfbox{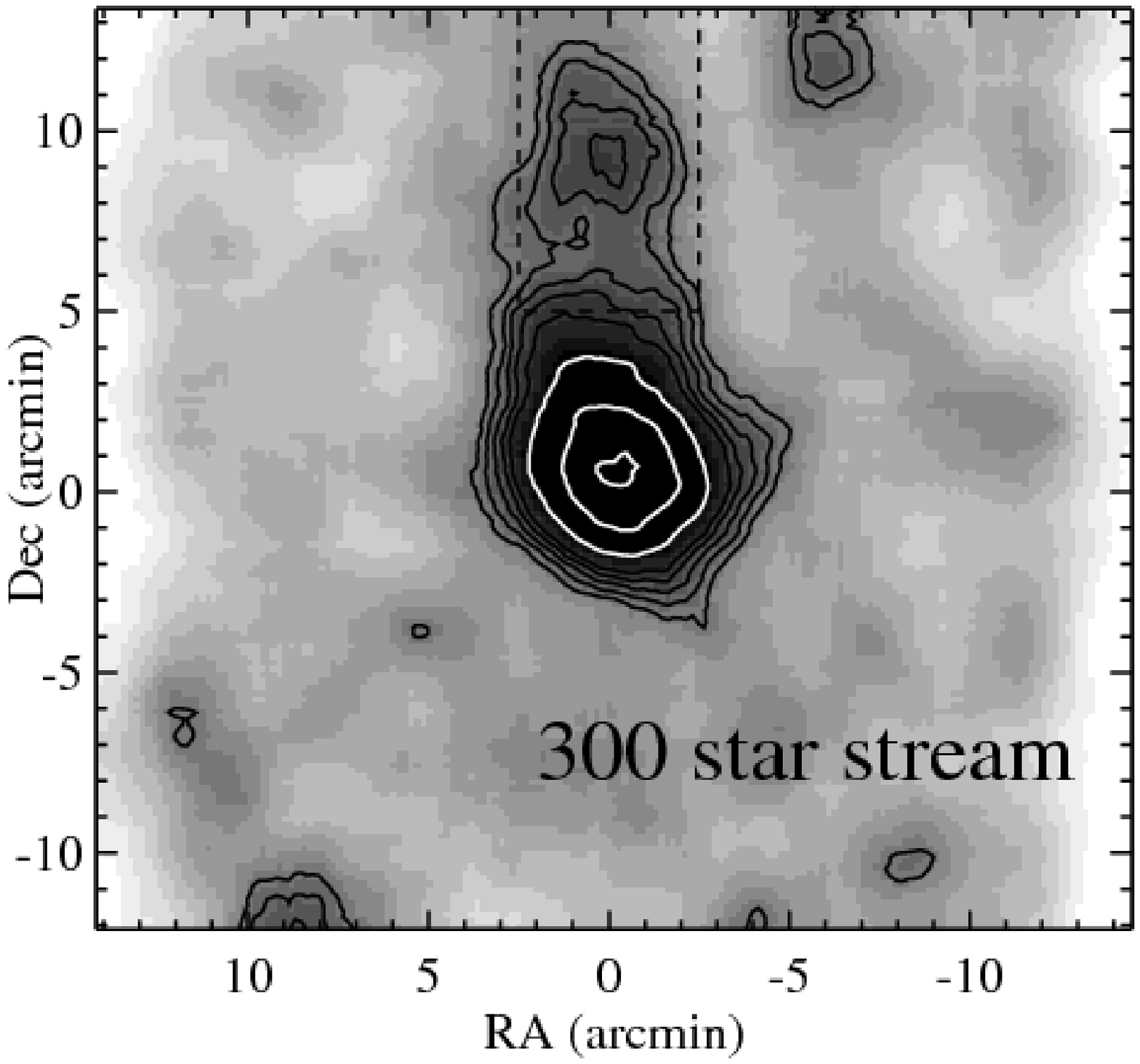}
}
\mbox{
\epsfysize=6.0cm \epsfbox{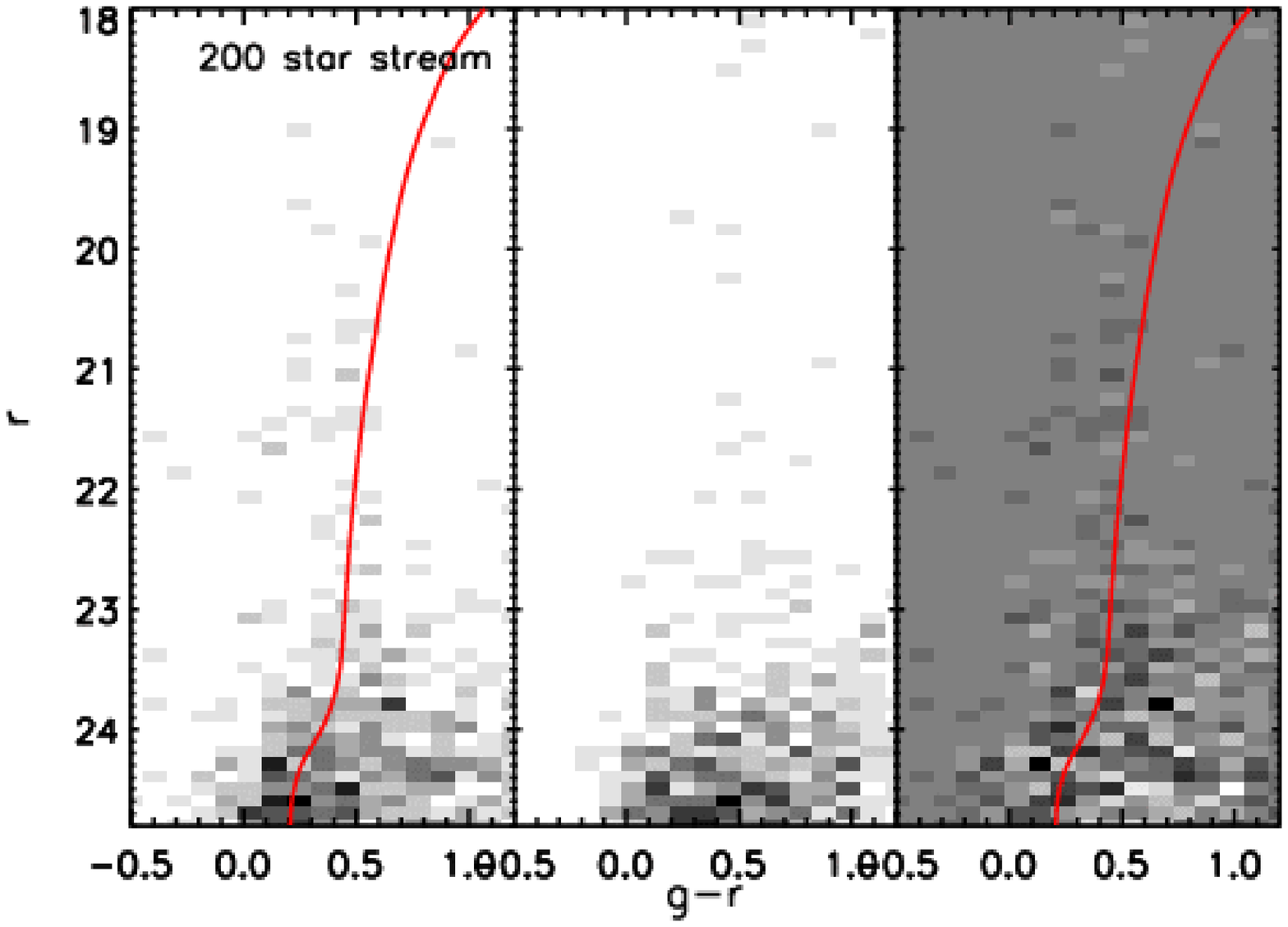}
\epsfysize=6.0cm \epsfbox{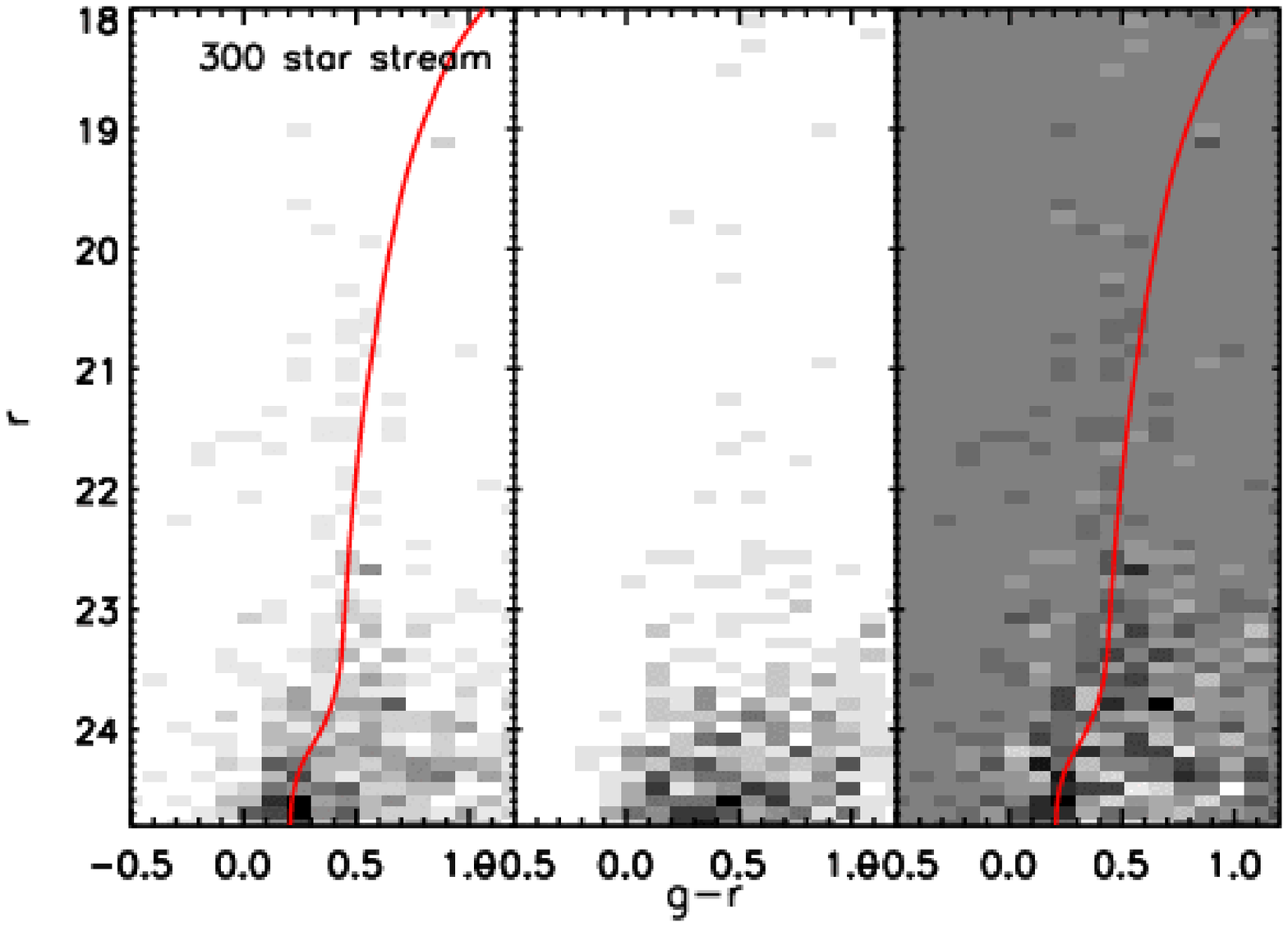}
}

\caption{ An illustration of our technique for implanting fake Leo~IV
'streams' into our field.  Top: We show our smoothed maps (using a 1
arcminute Gaussian) of Leo~IV after implanting streams with different
numbers of stars (see Table~\ref{tab:fakeresult}).  Note that the 200
and 300 star streams are clearly detected.  Bottom: The resulting Hess
diagrams for our 200 and 300 star scenarios.  Note that each has many
obvious red giant and BHB members, while the 300 star stream has the
beginnings of the main sequence.  We show the region we used for
extracting the 'stream' photometry in our maps in the top row of this
figure.
\label{fig:strinj}}
\end{center}
\end{figure*}

We inject both 'nuggets' -- Leo~IV-like stellar populations with an
exponential profile having $r_{half}=1.0$ arcmin -- and 'streams' --
with a Gaussian density profile in the right ascension direction with
$\sigma$=1.5 arcmin and a uniform distribution in the declination
direction over the Northern half of the field -- into our final Leo~IV
photometric catalog.  In Figures~\ref{fig:nuginj} and
\ref{fig:strinj}, we illustrate some results from our tests, along
with their properties in Table~\ref{tab:fakeresult}.

We show an example of a 35 star 'nugget' near what we believe is our
detection threshold in Figure~\ref{fig:nuginj}.  While this nugget has
a peak detection at 6.4-$\sigma$, it is not particularly different in
morphology or significance than the other random peaks in our Leo~IV
field.  This, however, changes when the resulting Hess diagram is
examined, showing several stars along the red giant branch --
something that is not apparent in the true overdensities in our field.
Taking this as our rough detection limit for compact remnants of
Leo~IV, we are sensitive to objects with a central surface brightness
of $\mu_{0,r}=27.9$ and $\mu_{0,g}=28.3$.

Turning towards our artificial 'streams', both the 200 and 300 star
streams are easily detectable in our smoothed maps (where we have used
a 1 arcminute smoothing scale to better pick out the thick streams --
Figure~\ref{fig:strinj}).  Further, the resulting Hess CMDs show a
considerable red giant and BHB presence, and the beginnings of the
main sequence for the 300 star case.  The analogous 100 star stream is
not convincingly detected.  We thus suggest that we are reliably
sensitive to streams with central surface brightness $\mu_{r}\sim29.6$
and $\mu_{g}\sim29.8$ (as measured along the center of the stream)
with a geometry and morphology roughly similar to that simulated.

The recent work of \citet{leoivleov} have presented tentative evidence
for a stream connecting Leo~IV and Leo~V with a surface brightness of
$\sim$32 mag arcsec$^{-2}$.  Unfortunately, despite our much deeper
data, we are unable to probe down to such faint surface brightness
limits due to the relatively small area of our single pointing.  

\subsection{Structure along the line of sight}\label{sec:los}

Because previous authors have suggested that the 'thick' red giant
branch in Leo~IV may be due to elongation along the line of sight, it
is worth searching for signs of such elongation in the width of the
BHB \citep[e.g.,][]{klessen03}.  To do this, we have created a BHB
star fiducial using data collected by J. Strader from SDSS
encompassing 41 horizontal branch stars from M3 and M13, corrected for
extinction and their relative distances.  Using these stars, a
third-order polynomial was fit with the IDL routine {\sc
robust\_poly\_fit}, and we used this polynomial to define our BHB
fiducial.  

For simplicity, we assume that all stars in our Megacam field within
the solid box in Figure~\ref{fig:CMD} are BHB stars belonging to
Leo~IV.  Placing the BHB fiducial at our assumed distance to Leo~IV,
$(m-M)=20.94$, leads to a visually excellent match.  However, since we
are interested in the scatter about this fiducial to put constraints
on the width of Leo~IV, rather than the absolute distance to it, we
adjust the fiducial BHB sequence so that the average $r$ magnitude
deviation of the Leo~IV BHB stars against the fiducial is $\sim$0.0
(this adjustment was 0.06 mag, affirming the distance measurement of
\citet{Moretti09}).  The resulting root mean square deviation of the
Leo~IV BHB stars about the fiducial is $\sim$0.2 mag, which
corresponds to a deviation of $\sim$15 kpc at the distance of Leo~IV.
While this limit is comparable to the quoted difference in distance
between Leo~IV and Leo~V \citep{leov}, it is also roughly the
measurement limit achievable using the spread in BHB mags as a measure
of line-of-sight depth.  First, there are known RR~Lyrae variables
among the BHB stars in Leo~IV \citep{Moretti09}, which we have
observed at a random phase.  Second, for a spread in metallicity of
$\sigma_{[Fe/H]}=0.75$ in Leo~IV \citep{Kirby08}, one would expect a
natural spread in BHB star magnitudes of $\sim$0.2 mag
\citep{sandage90,olszewski96}.  We therefore conclude that our data,
while consistent with no elongation along the line of sight, can not
put strong constraints on the depth of Leo~IV.

\section{Stellar population}\label{sec:starform}

There is spectroscopic evidence that Leo~IV has a large metallicity
spread, and according to \citet{Kirby08} it has the largest
metallicity spread seen among the new dwarfs ($\langle [Fe/H]
\rangle$=$-$2.58 with a spread of $\sigma_{[Fe/H]}$=0.75).
Additionally, \citet{Belokurov07} have hinted that Leo~IV has a
'thick' red giant branch, indicating a spread in metallicity, and this
appears to be the case (at least superficially) in the CMD presented
in the current work (Figure~\ref{fig:CMD}).  Also intriguing is the
population of blue plume stars pointed out in Figure~\ref{fig:CMD} and
\S~\ref{sec:fincat}.  In this section, we first perform a CMD-fitting
analysis of the stellar population of Leo~IV with StarFISH, and then
go on to look at the blue plume population in detail.  We end the
section by combining the structural properties found in
\S~\ref{sec:structparams} with the stellar population determined in
the current section to calculate the total absolute magnitude of
Leo~IV.

\subsection{Star Formation History via CMD-fitting}\label{sec:starfish}

Here we apply the CMD-fitting package StarFISH \citep{starfish} to our
Leo~IV photometry within the half light radius to determine its SFH
and metallicity evolution.  As discussed in previous works, StarFISH
uses theoretical isochrones (we use those of Girardi et al. 2004,
although any may be used) to construct artificial CMDs with different
combinations of distance, age, and metallicity.  Once convolved with
the observed photometric errors and completeness (using the artificial
star tests of \S~\ref{sec:observations}), these theoretical CMDs are
converted into realistic model CMDs which can be directly compared to
the data on a pixel-to-pixel basis, after binning each into Hess
diagrams.  We use the Poisson statistic of \citet{match} as our fit
statistic.  The best fitting linear combination of model CMDs is
determined through a modification of the standard AMOEBA algorithm
\citep{Press88,starfish}.  Several steps are taken to determine the
uncertainties in StarFISH fits, which are discussed in detail in
previous work -- see \citet{Harrislmc,leot}.

Our StarFISH analysis is similar to that of \citet{sandherc}.  We
include isochrones with [Fe/H]=-2.3,-1.7 and -1.3 and ages between
$\sim$10 Myr and $\sim$15 Gyr.  Age bins of width $\Delta$log(t)=0.4
dex were adopted, except for the two oldest age bins centered at
$\sim$10 Gyr and $\sim$14 Gyr, where the binning was
$\Delta$log(t)=0.3 dex.  A 'background/foreground' CMD, created by
taking all stars outside an elliptical radius of 12 arcminutes (with
ellipticity of $\epsilon$=0.05, as in our best-fitting exponential
profile, see Table~\ref{table:paramfits}), was simultaneously fit
along with our input stellar populations in order to correct for
contamination by unresolved galaxies and foregound stars in our Leo~IV
CMD.

Stars with magnitudes $18.0<r<24.75$ (corresponding roughly to our
50\% completeness limit) and $-0.50 < g-r < 1.15$ were fit.  After
some experimentation, we used a Hess diagram bin size of 0.15 mag in
magnitude and 0.15 mag in color.  We assume a binary fraction of 0.5
and a Salpeter initial mass function.  Because we correct our stellar
catalog for Galactic extinction with the dust maps of
\citet{Schlegel98}, we do not allow the mean extinction of our model
CMDs to vary.  As in the rest of this work, we chose to fix the
distance modulus in the code to $(m-M)$=20.94 mag, although our
results are robust with respect to this assumption, as we discuss
below.

\begin{figure*}
\begin{center}
\mbox{
\epsfysize=7.0cm \epsfbox{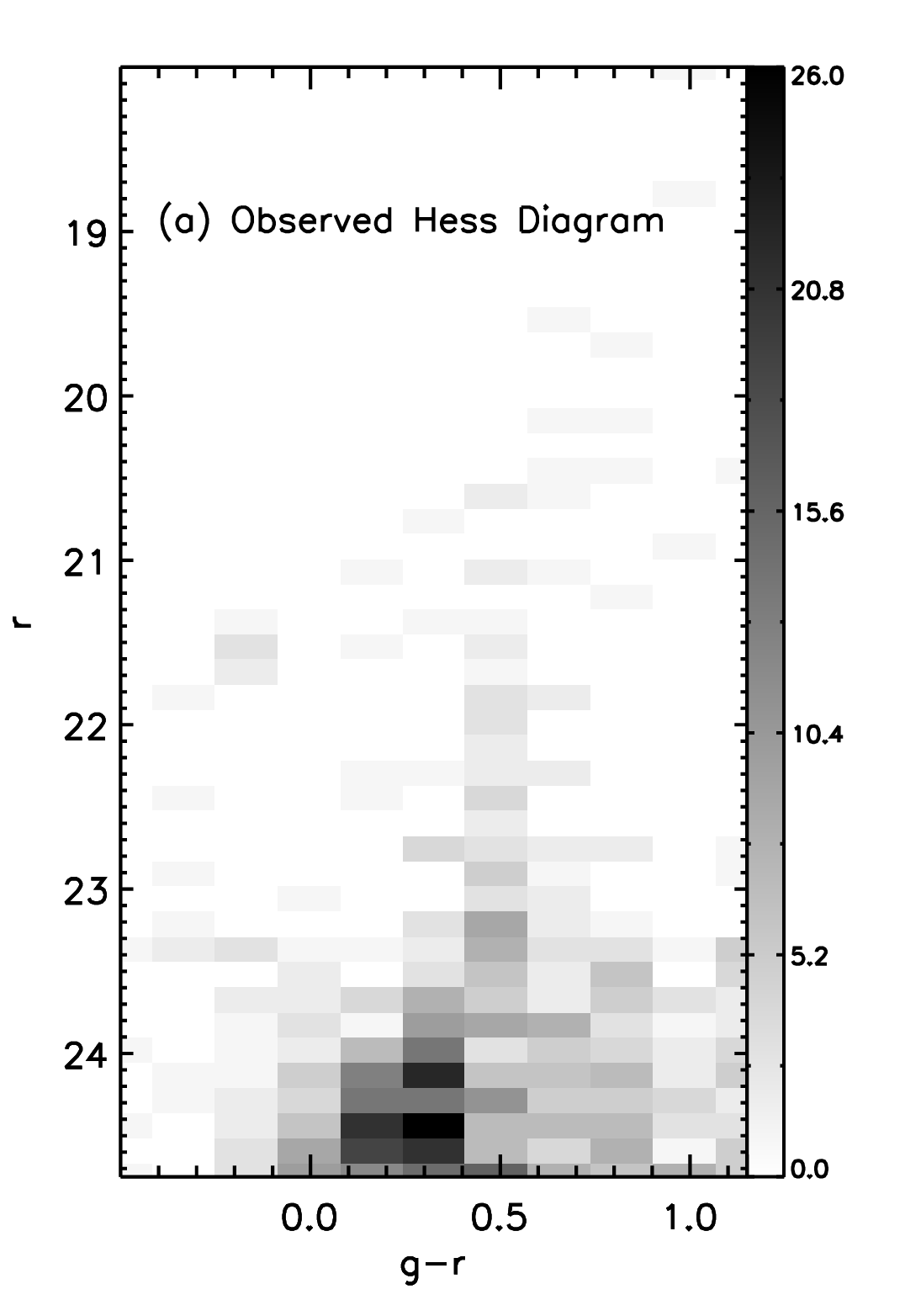}
\epsfysize=7.0cm \epsfbox{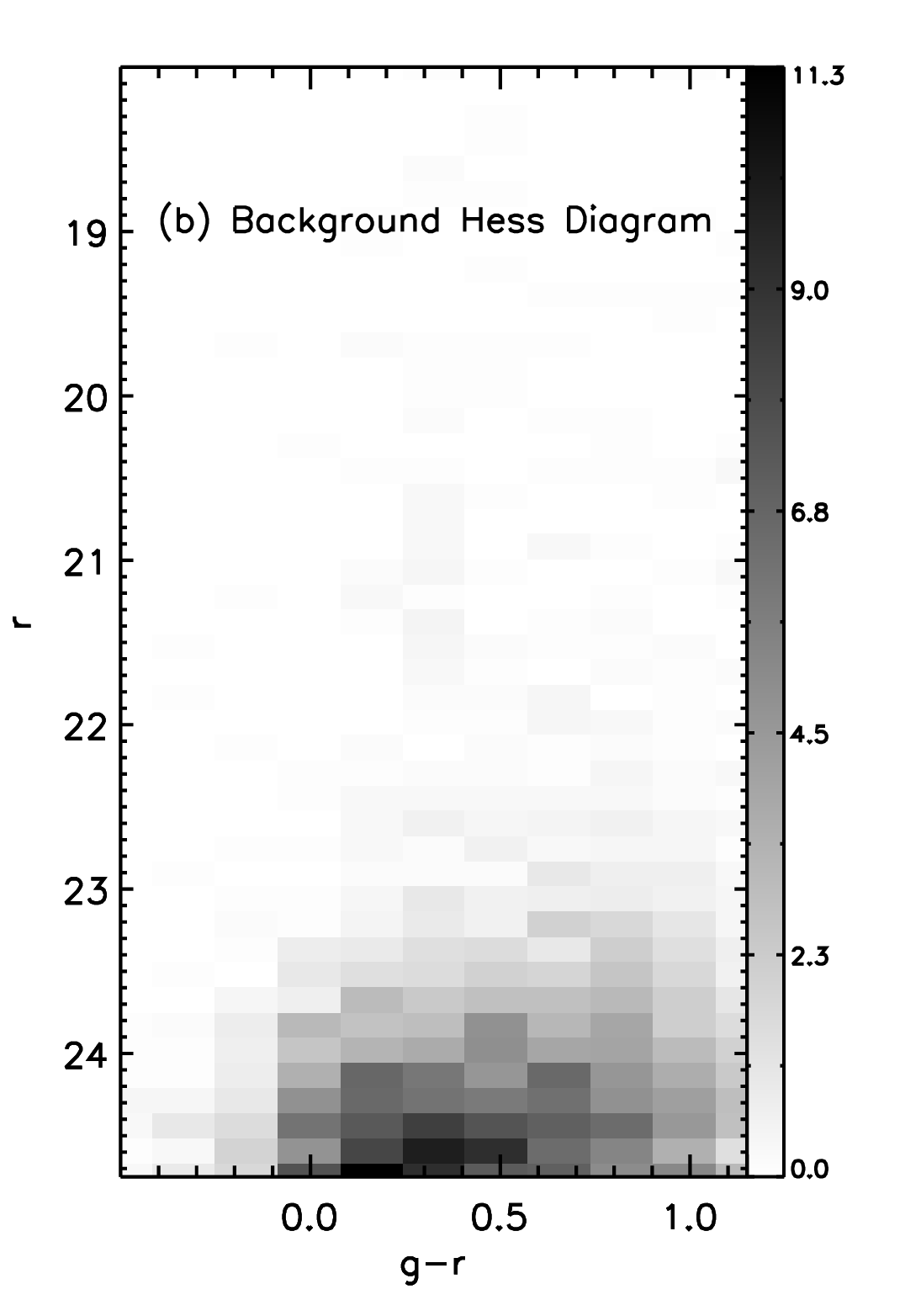}
\epsfysize=7.0cm \epsfbox{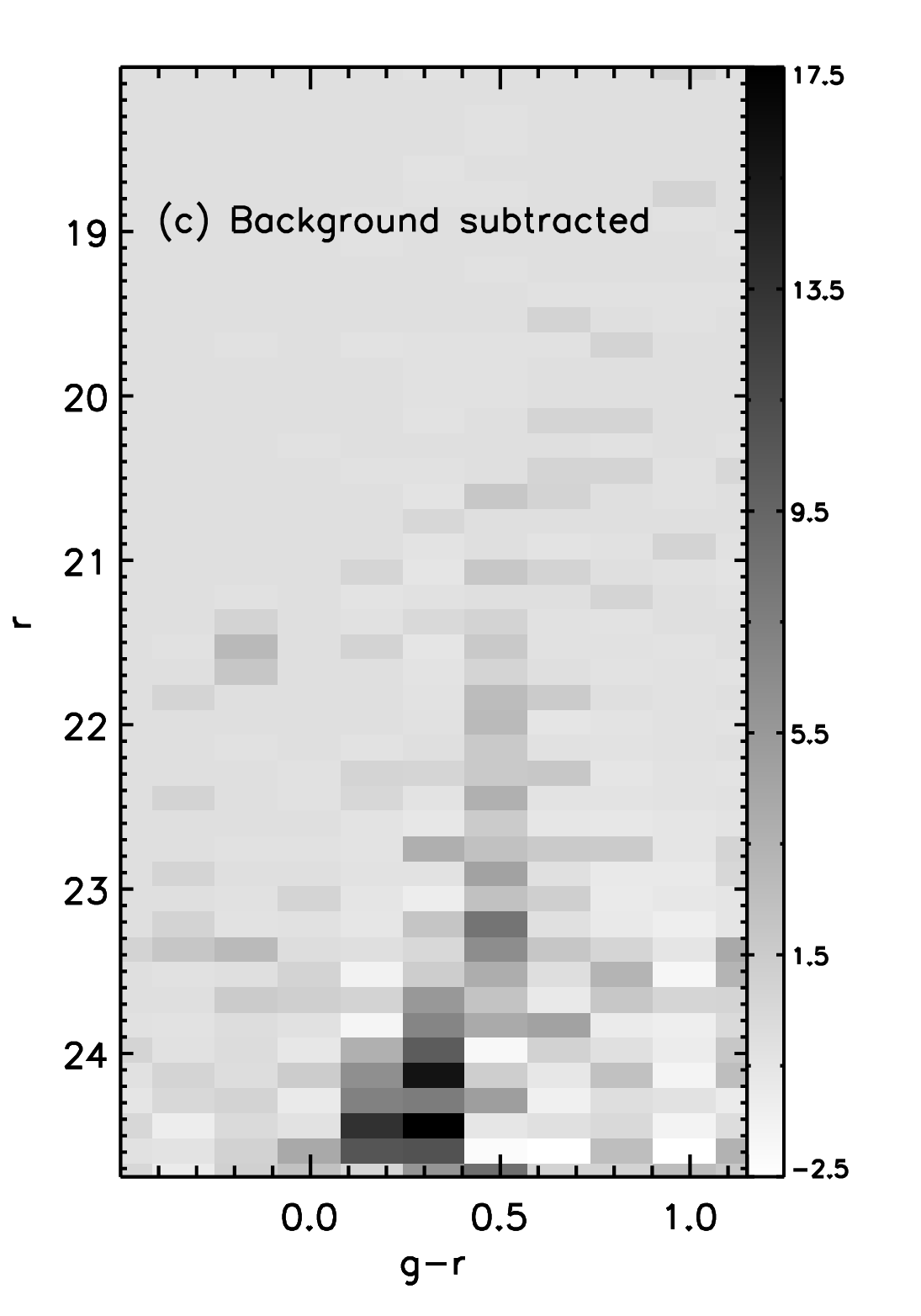}
}
\mbox{
\epsfysize=7.0cm \epsfbox{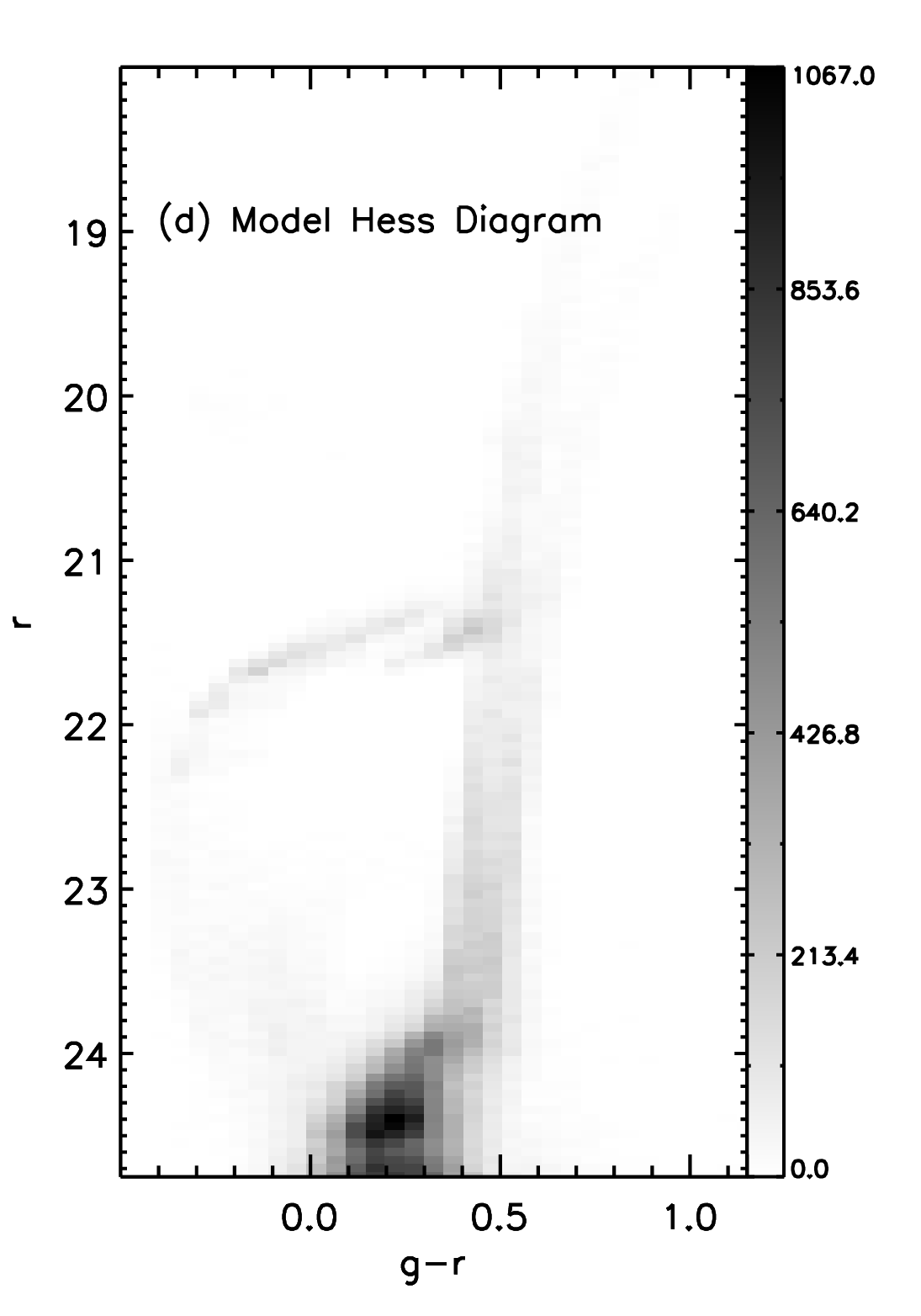}
\epsfysize=7.0cm \epsfbox{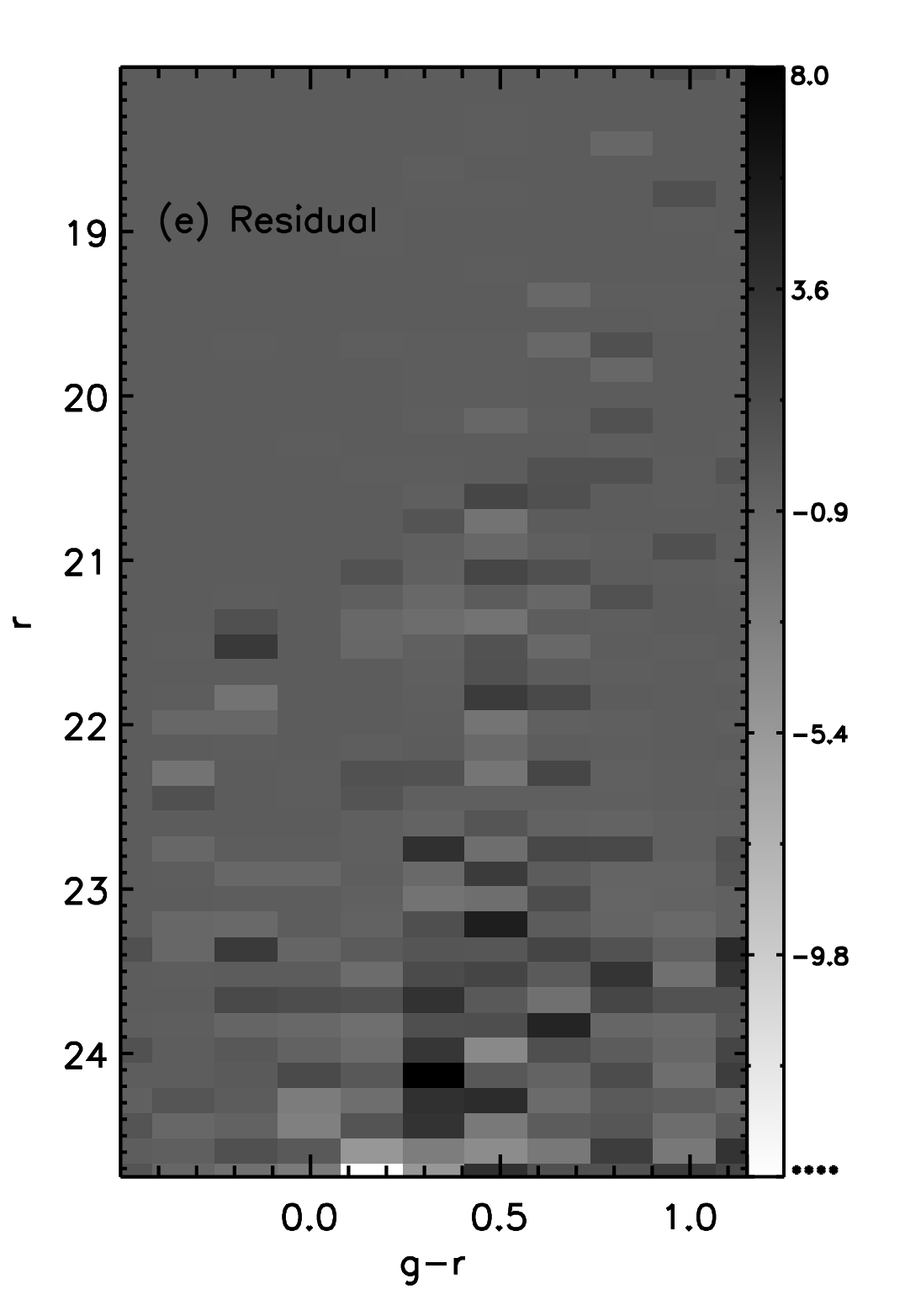}
\epsfysize=7.0cm \epsfbox{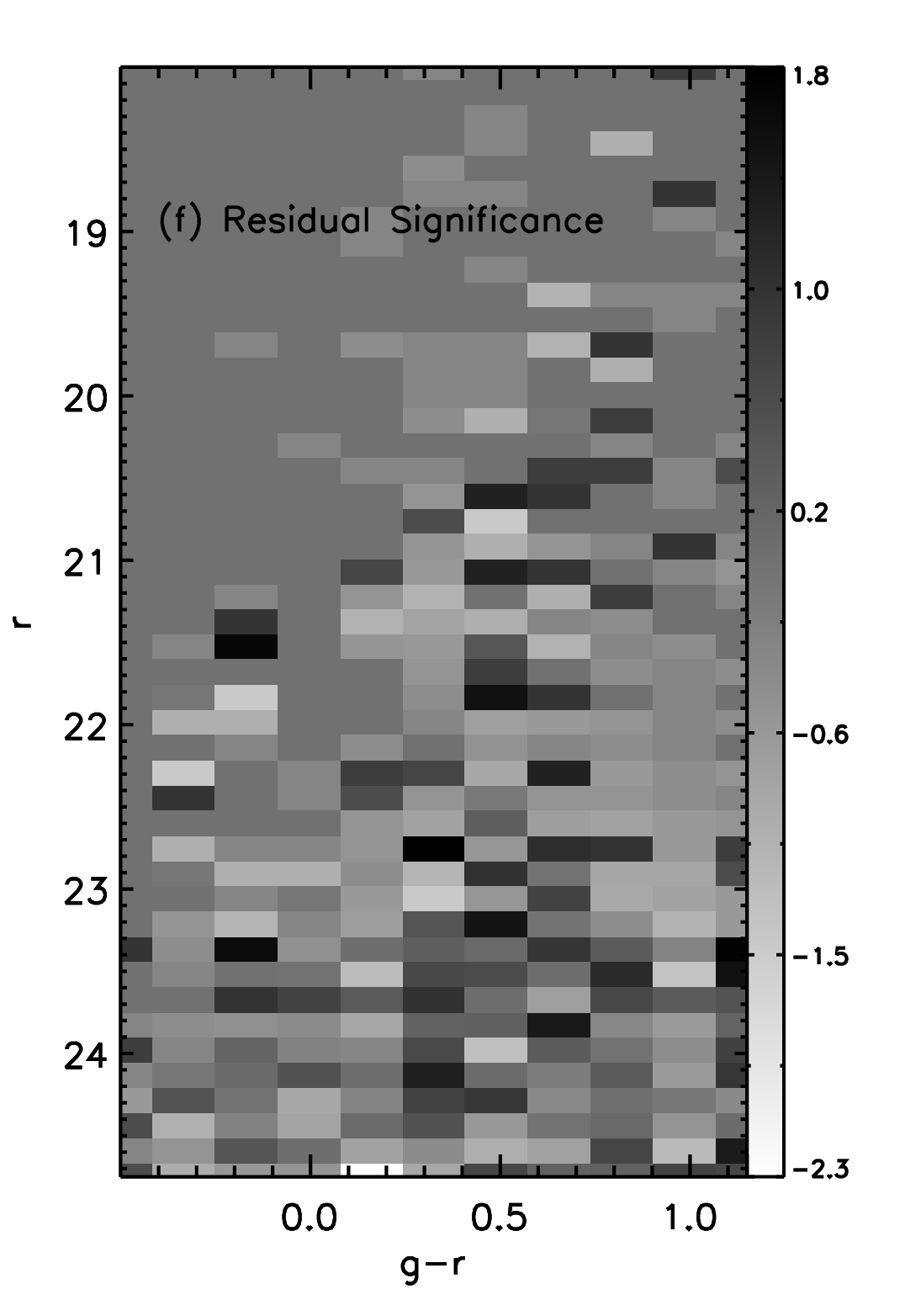}
}

\caption{ Comparison of the data with our best StarFISH model fit.
Pixel bins are 0.15 mag along the color and magnitude axis (except for
panel (d), where the bins are 0.05 mag), as used in our StarFISH fits.
The gray scale for each Hess diagram is in units of stars per pixel,
except for panel (f), which has been scaled by the uncertainty
associated with each pixel.  (a) The observed Hess diagram of Leo~IV
within the half light radius.  (b) The background Hess diagram, which
was fit along with the theoretical isochrones.  (c) The
background-subtracted Hess diagram.  (d) The best model CMD derived by
StarFISH (with no background component), in bins of 0.05 mag in order
to show the details of the model.  (e) Residuals after subtracting a
random realization of the StarFISH model from the data.  (f) The
residual from (e), scaled by the expected Poisson scatter in the bin.
The most significant residuals are associated with the mismatch of the
BHB between model and data.
\label{fig:sfh_model}}
\end{center}
\end{figure*}

We show the best-fit StarFISH solution in Figure~\ref{fig:sfh}, along
with a comparison of the best model CMD with that of Leo~IV in
Figure~\ref{fig:sfh_model}.  Leo~IV is consistent with a single
stellar population with an age of $\sim$14 Gyr and a [Fe/H]=-2.3,
although the error bars and upper limits indicate that there is
latitude for both a small, young stellar population and a mix of
metallicities at older ages.  Thus, despite the visual impression of a
'thick' giant branch, our analysis does not require a metallicity
spread to match the oberved CMD.  Thorough spectroscopy of all stars
in the red giant region will be required to determine Leo~IV
membership and to properly quantify the metallicity spread.  This
result is robust to small changes in the distance modulus.  If we
alter the input distance modulus from $(m-M)$=20.94 to $(m-M)$=20.84,
than a larger fraction of the best-fitting SFH comes from the
[Fe/H]=-1.3 bin, although the [Fe/H]=-2.3 bin still dominates.
Likewise, a distance modulus of $(m-M)$=21.04 yields an old,
[Fe/H]=-2.3 stellar population with even less from the [Fe/H]=-1.7 and
-1.3 bins.


The model and observed CMDs match remarkably well
(Figure~\ref{fig:sfh_model}) given that there are known mismatches
between the theoretical isochrones and empirical, single population
CMDs \citep[e.g.,][]{Girardi04}.  Also, the available models do not
span the metallicity range that is apparent in the new MW dwarfs;
\citet{Kirby08} found $\langle [Fe/H] \rangle=-2.58$ with an intrinsic
scatter of $\sigma_{[Fe/H]}=0.75$ in Leo~IV, while the Girardi
isochrone set reaches down to [Fe/H]=-2.3.  There is a slight mismatch
in the BHB between model and data, with the best-fitting model CMD
having a BHB which is $\sim$0.1-0.2 mag brighter than that observed
(this is a factor of $\sim$2 larger than the small BHB magnitude
correction we made in \S~\ref{sec:los}).  This could be due to a
slight error in our assumed distance modulus or a true stellar
population that is even more metal poor than the most metal-poor model
in the Girardi isochrone set, which we know to be the case from
\citet{Kirby08}.  Nonetheless, our basic finding that Leo~IV is
predominantly old ($\sim$14 Gyr) and metal poor is unaffected.

\subsection{Evidence for Young Stars}\label{sec:BP}

It is well known that dwarf spheroidals harbor a population of blue
stragglers -- a hot, blue extension of objects which lie along the
normal main sequence \citep[e.g.,][]{Mateo95}.  Because the densities
in dwarfs do not reach that necessary to produce collisional binaries
(as they can in the cores of globular clusters), it is likely that
they are primordial binary systems, coeval with the bulk of stars in
the dwarf.  Unfortunately, their position along the main sequence
makes it difficult to disentangle blue straggler stars from young main
sequence stars in the MW's dwarf spheroidals, which has been a
continuous source of ambiguity \citep[e.g.,][among many
others]{Mateo95,Mapelli07,Mapelli09}. It is very difficult to exclude
the possibility that some of the blue plume stars in any given dwarf
are actually young main sequence objects.

We now articulate two arguments in support of the hypothesis that at
least some of the stars in the blue plume of Leo~IV are young.  First,
the blue plume stars appear to be segregated within the body of
Leo~IV.  We plot the position of the 'high probability' blue plume
stars with low background/foreground contamination, as identified
within the dotted box in Figure~\ref{fig:CMD}, onto our smoothed map
of Leo~IV (Figure~\ref{fig:showstars}).  All of the selected
high-probability blue plume stars within the body of Leo~IV are on one
side.  If Leo~IV is assumed to be spherically symmetric, the chances
of all seven being on the same half of the galaxy are $\sim(1/2^{7})$
or less than 1 percent.  This segregation would be difficult to
explain if all of these objects were blue stragglers, which would
presumably have the same distribution as the galaxy as a whole.  We
understand that this {\it a posteriori} argument is insufficient on
its own, but it should be considered in the context of the high
normalized blue plume fraction, which we now discuss.

Our second argument in favor of a young population of stars in Leo~IV
stems from the high blue plume frequency normalized by the BHB star
counts, following the work of \citet{Momany07}.  Briefly,
\citet{Momany07} sought to explore the ambiguity between young main
sequence stars and genuine blue straggler stars in a sample of MW
dwarf galaxies by calculating the number of total blue plume objects
with respect to a reference stellar population -- the BHB.  The basic
result of their work, whose data was kindly provided by Y. Momany, was
that those dwarf spheroidals that do not have a true, young stellar
component have a blue plume fraction that follows a relatively well
defined $M_{V}$ vs log($N_{BP}/N_{BHB}$) anti-correlation, where the
blue plume consists of only blue straggler stars
(Figure~\ref{fig:bpfreq}).  The physical origin of this
anti-correlation is unclear \citep[see, however, ][for a plausible
explanation for the anti-correlation in globular clusters]{Davies04},
although it is seen in both open clusters \citep{deMarchi08} and
globular clusters \citep{Piotto04}.  Carina, the one dwarf galaxy in
their sample which does have a known, recent bout of star formation
\cite[$\sim$1-3 Gyr;][]{hurleykeller98,monelli03}, was shown to have
an excess of blue plume stars with respect to the aforementioned
relation (shown as a lower limit in Figure~\ref{fig:bpfreq}, due to
difficulties in accounting for blue stragglers associated with the
older and fainter main sequence turn-off in that system), pointing to
the fact that Carina's blue plume was populated with young main
sequence stars in addition to a standard blue straggler population.
More recently, \citet{Martin08} also found a high blue plume frequency
in Canes Venatici I, which we also show in Figure~\ref{fig:bpfreq}.
\citet{Martin08} used the combination of spatial segregation and blue
plume frequency to argue that Canes Venatici I harbors a young stellar
component.

We take the ratio of blue plume to BHB stars as identified in
Figure~\ref{fig:CMD} by the dashed and solid regions, respectively,
utilizing those stars within the half light radius of Leo IV and
making background and completness corrections. We calculate a blue
plume frequency of log($N_{BP}/N_{BHB}$)=$0.56\pm0.13$ for Leo~IV and
plot this ratio along with those of the MW dwarf spheroidals just
discussed in Figure~\ref{fig:bpfreq}.  As can be seen, Leo~IV lies off
the standard $M_{V}$ vs log($N_{BP}/N_{BHB}$) anti-correlation just as
Canes Venatici I and Carina do.  Leo~IV lies $2-\sigma$ away from the
linear relation fit by \citet{Momany07} for the non-star forming
dwarfs.  

Taken together, the segregation of blue plume stars along with the
high blue plume to BHB fraction points to at least {\it some} of the
stars being young main sequence objects.  We know that at least one of
these blue plume stars is a blue straggler, due to the discovery of
one SX Phoenicis variable in Leo~IV \citep{Moretti09}.  This is only
the third of the new dwarf spheroidals (excluding Leo~T, which appears
to be a transition object) that harbor a young stellar population --
the others being the previously discussed Canes Venatici I
\citep{Martin08} and Ursa Major II \citep{sdsssfh}.  We note that our
evidence for young stars is similar to that presented for Canes
Venatici I, for which it was also found that the blue plume population
is segregated and blue straggler frequency is high \citep{Martin08}.
We measure the luminosity of the young stars in \S~\ref{sec:absmag}
and discuss the implications of Leo~IV's young stellar component in
\S~\ref{sec:discuss}.

\subsection{Absolute Magnitude}\label{sec:absmag}

As pointed out by \citet{sdssstruct}, measuring the total magnitudes
of the new MW satellites is difficult due to the small number of stars
at detectable levels.  We account for this 'CMD shot noise' by
mimicking our measurement of the total magnitude of Hercules
\citep{sandherc}, which borrowed heavily from the original
\citet{sdssstruct} analysis.

We take the best SFH solution presented in \S~\ref{sec:starfish}, and
create a well-populated CMD (of $\sim$200000 stars) incorporating our
photometric completeness and uncertainties, using the {\it repop}
program within the StarFISH software suite.  We drew one thousand
random realizations of the Leo~IV CMD with an identical number of
stars as we found for our exponential profile fit, and determined the
'observed' magnitude of each realization above our 90\% completeness
limit.  We then accounted for those stars below our 90\% completeness
level by using luminosity function corrections derived from
\citet{Girardi04}, using an isochrone with a 15 Gyr age and
[Fe/H]=-2.3.  We take the median value of our one thousand random
realizations as our measure of the absolute magnitude and its standard
deviation as our uncertainty (Table~\ref{table:paramfits}).  To
convert from $M_{r}$ magnitudes to $M_{V}$ magnitudes we use
$V-r=0.16$ \citep{Walsh08}.


We find $M_{V}=-5.5\pm0.3$ and a central surface brightness, assuming
our exponential profile fit, of $\mu_{0,V}=27.2\pm0.3$.  Both our
total absolute magnitude and central surface brightness measurement
agrees to within $1-\sigma$ with the measurement of
\citet{sdssstruct}, which used only SDSS data.

We use a similar technique for determining the approximate luminosity
of Leo~IV's young stellar population only.  We draw an equivalent
number stars from our well-populated StarFISH CMD as in Leo~IV within
our high probability blue plume box (see Figure~\ref{fig:CMD} and
\S~\ref{sec:BP}), and corrected for the luminosity of stars outside
this region using a luminosity function derived from an isochrone with
an age of 1.6 Gyr and [Fe/H]=-1.3 \citep{Girardi04}.  We again draw
1000 random realizations to determine our uncertainties.  We find that
the young stellar population has $M_{V}=-2.1\pm0.5$, or roughly
$\sim$5\% of the satellite's total luminosity or $\sim$2\% of its
stellar mass.  This magnitude and the resulting fraction of the young
stellar populations luminosity should be taken with caution given our
assumptions and the small number of stars involved.

\section{Discussion \& Conclusions}\label{sec:discuss}

In this work we have presented deep imaging of the Leo~IV MW satellite
with Megacam on the MMT and study this galaxy's structure and SFH.  In
particular, we assess reports in the literature concerning both its
stellar population and its possible association with the nearby
satellite, Leo~V.

Leo~IV's SFH is dominated by an old ($>12$ Gyr), metal poor
($[Fe/H]\lesssim-2.0$) stellar population, although we uncover
evidence for a young sprinkling of star formation 1-2 Gyrs ago.  Our
best-fit StarFISH results indicate that a single metal poor population
dominates, although the data is also compatible with a spread in
metallicities.  The old population is consistent with the emerging
picture that the faintest MW satellites are 'reionization fossils'
\citep[e.g.~][]{Ricotti05,Gnedin06}, who formed their stars before
reionization and then lost most of their baryons due to
photoevaporation.  The apparent sprinkling of young stars begs the
question of what has enabled Leo~IV to continue forming stars at a low
level.  There is no sign of HI in Leo~IV, with an upper limit of 609
$M_{\odot}$ \citep{Grcevich09}, although we note that this limit is
still a factor of $\sim$2 larger than the stellar mass associated with
the young stellar population studied in \S~\ref{sec:BP}.

One possible mechanism for late gas accretion, and subsequent star
formation, among the faint MW satellites was recently discussed by
\citet{Ricotti09} to help explain the complex SFH and gas content of
Leo~T \citep{leot,leotgas}.  In this scenario, the smallest halos stop
accreting gas after reionization as expected, but as their temperature
decreases and dark matter concentration increases with decreasing
redshift they are again able to accrete gas from the intergalactic
medium at late times, assuming they themselves are not accreted by
their parent halo until $z\lesssim1-2$.  This can lead to a bimodality
in the SFH of the satellite, with both a $>12$ Gyr population and one
that is $<10$ Gyr, as we see in Leo~IV.  One stringent requirement of
this model is that the satellite can not have been accreted by its
host halo until $z\lesssim1-2$ (and thus not exposed to tidal stirring
and ram pressure stripping until late times, allowing the satellite to
retain its newly accreted gas).  Future proper motion studies will be
able to test if Leo~IV is compatible with this late gas accretion
model.  Another prediction of this model is the possible existence of
gas-rich minihaloes that never formed stars, but could serve as fuel
for star formation if they encountered one of the luminous dwarfs.
More detailed study of this late gas accretion mechanism will be
necessary to understand the possible diversity of SFHs in the faint MW
satellites.

Additionally, we note that if the apparent segregation of young stars
in Leo~IV is real, then it is not an isolated case among the MW
satellites.  As has been mentioned in \S~\ref{sec:BP},
\citet{Martin08} noted that Canes~Venatici I has a compact star
forming region clearly offset from the galaxy as a whole, with an age
of 1-2 Gyrs, similar to Leo~IV.  Additionally, Fornax has several
compact clumps and shells that house young stellar populations roughly
$\sim$1.4 Gyrs old \citep{coleman04,Coleman05,Olszewski06}.  It has
been suggested that these could have been the result of a collision
between Fornax and a low-mass halo, which was possibly gas-rich.
\citet{pennarubia09} investigated the disruption of star clusters in
triaxial, dwarf-sized halos and found that segregated structures can
persist depending on the orbital properties of the cluster, providing
yet another viable mechanism.  More work is needed to distinguish
between all of the above scenarios and to properly model the emerging
diversity of SFHs among the new, faint MW satellites.

Structurally, Leo~IV appears to be very round, with $\epsilon \lesssim
0.23$ (at the 68\% confidence limit) and a half light radius ($\sim
130$ pc) which is typical of the new MW satellites.  An exhaustive
search for signs of extended structure in the plane of the sky has
ruled out any associated streams with surface brightnesses of
$\mu_{r}\lesssim29.6$.  The extent of Leo~IV along the line of sight
is less than $\sim$15 kpc, a limit that will be difficult to improve
upon given the inherent limitations of using the spread in BHB
magnitudes to measure depth.  We find no evidence for structural
anomalies or tidal disruption in Leo~IV.  We do not have the
combination of image depth and area necessary to confirm the stellar
bridge, with a surface brightness of 32 mag arcsec$^{-2}$, recently
reported in between Leo~IV and Leo~V \citep{leoivleov}.  Indeed, Leo~V
is almost certainly disrupting, as discussed by \citet{leovspec}, due
to the presence of two member stars $\sim$13 arcminutes ($>13 r_{h}$)
from Leo~V's center along the line connecting the putative
Leo~IV/Leo~V system.  The nature of Leo~V is still very ambiguous,
with the kinematic data being consistent with it being dark matter
free -- suggesting that perhaps Leo~V is an evaporating star cluster
\citep{leovspec}.  It is thus critical to obtain yet deeper data on
these two systems and the region separating them to uncover their true
nature.

The probable detection of a small population of young stars
illustrates once again that it is crucial to obtain deep and wide
field follow up for all of the newly detected MW satellites.  Every
new object has a surprise or two in store upon closer inspection.

\section{Acknowledgements}

We are grateful to the referee, Nicolas Martin, for his constructive
report.  Many thanks to Maureen Conroy, Nathalie Martimbeau, Brian
McLeod and the whole Megacam team for the timely help in reducing our
Leo~IV data.  DJS is grateful to Jay Strader for his excellent
scientific advice, and for providing his co-calibrated M3 \& M13 blue
horizontal branch sequence.  We are grateful to Evan Kirby, Josh Simon
and Marla Geha for providing their kinematic and metallicity data on
Leo IV.  Yazan Momany kindly provided his blue plume frequency data.
DJS is grateful to Matthew Walker and Nelson Caldwell for providing a
careful reading of the paper, along with useful comments.  EO was
partially supported by NSF grant AST-0807498.  DZ
acknowledges support from NASA LTSA award NNG05GE82G and NSF grant
AST-0307492.

\section{Appendix}

In this Appendix, we determine how well the maximum likelihood
technique presented in \S~\ref{sec:structparams} can measure the
structural parameters of a dwarf with a comparable number of stars as
Leo~IV. We create mock models of Leo~IV-like systems having an
exponential profile with $r_{h}$=3.0 arcmin and 400 stars in the input
catalog (all on a footprint with the same area as our MMT pointing),
while systematically varying the ellipticity and position angle. A
uniform background of 4.5 stars per square arcminute is randomly
scattered across the field to mimic the actual observations.  We
utilize the same algorithm as in \S~\ref{sec:structparams}, with 1000
bootstrap resamples to determine our uncertainties.

We can recover the ellipticity of our mock dwarf galaxies remarkably
well, as can be seen in Figure~\ref{fig:ellinout}.  Here we show our
results on the recovered ellipticity as a function of input model
ellipticity, between $\epsilon_{input}=$0.05 and 0.85.  The data
points are the median ellipticity found for the one thousand bootstrap
resamples for each model, and the error bars encompass 68\% of the
resamples around that median.  Note that for models with an input
ellipticity of $\epsilon \lesssim 0.25$ it is difficult for our
algorithm to converge on the correct ellipticity value.  In this
regime, we present the measured ellipticity as an upper limit (see
\S~\ref{sec:structparams}).  In this case, we systematically
overpredict the ellipticity with large error bars, and thus only quote
upper limits.  Larger values of the ellipticity are well measured.

\begin{figure*}
\begin{center}
\mbox{
\epsfysize=4.0cm \epsfbox{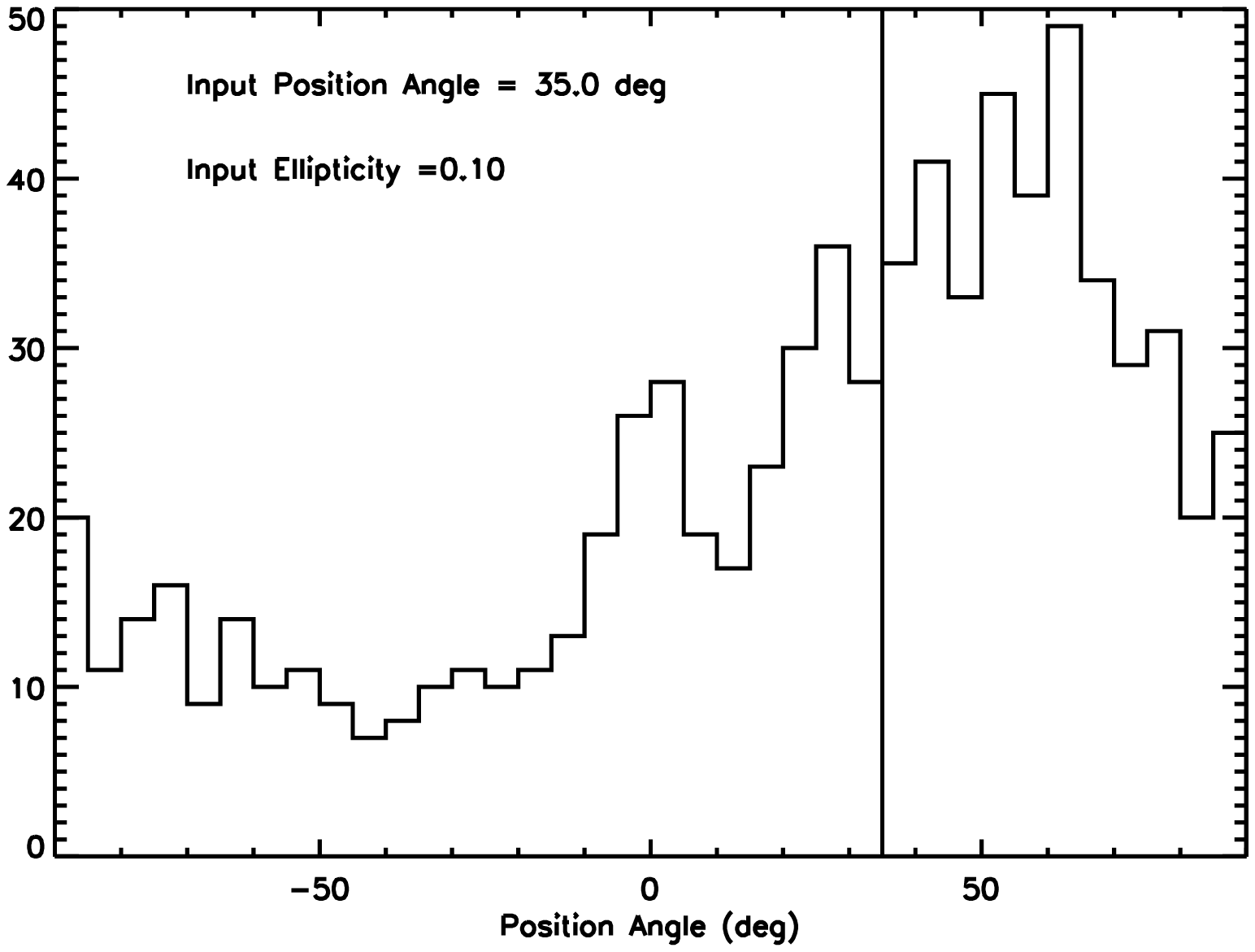}
\epsfysize=4.0cm \epsfbox{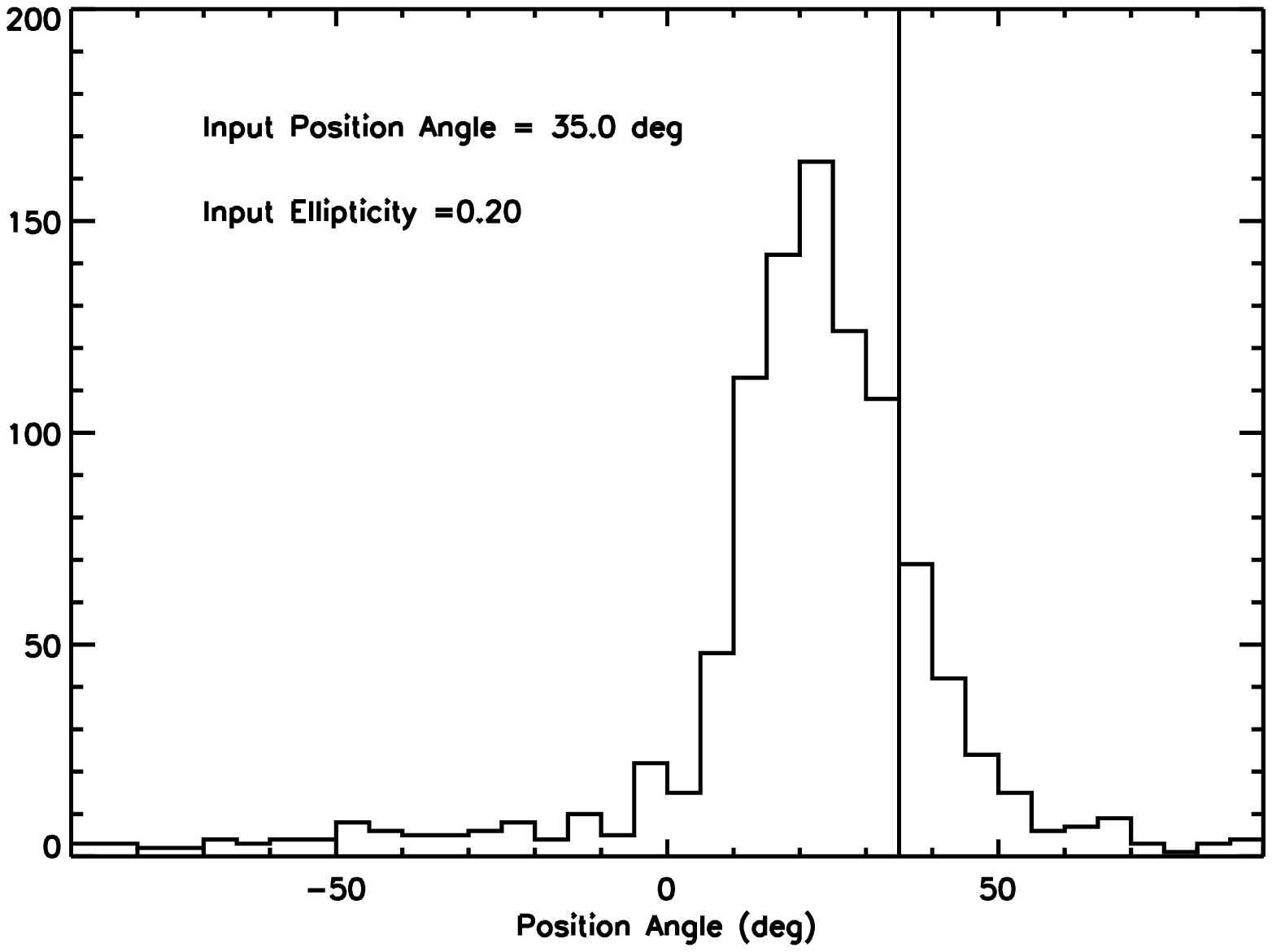}
\epsfysize=4.0cm \epsfbox{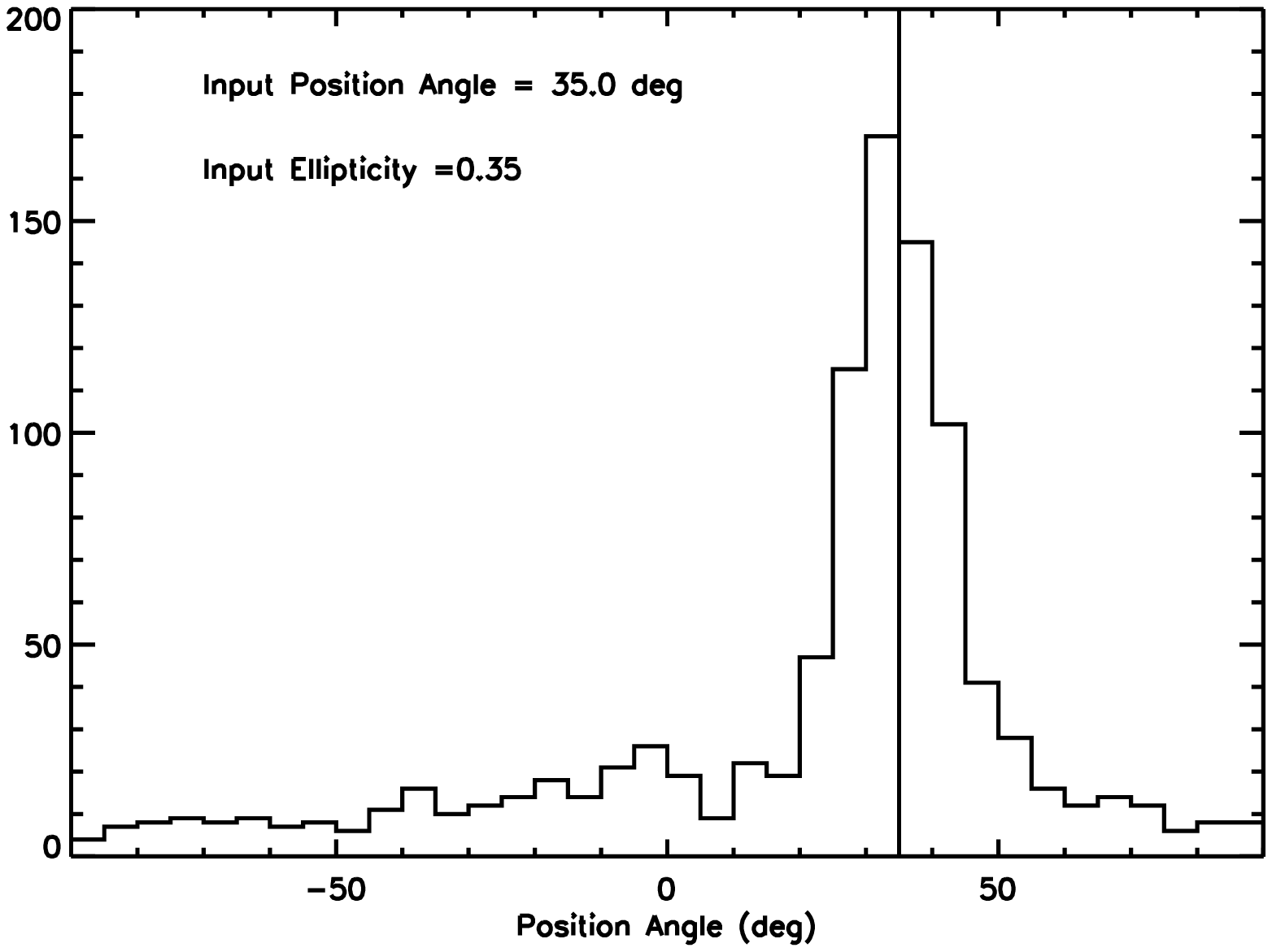}
}
\mbox{
\epsfysize=4.0cm \epsfbox{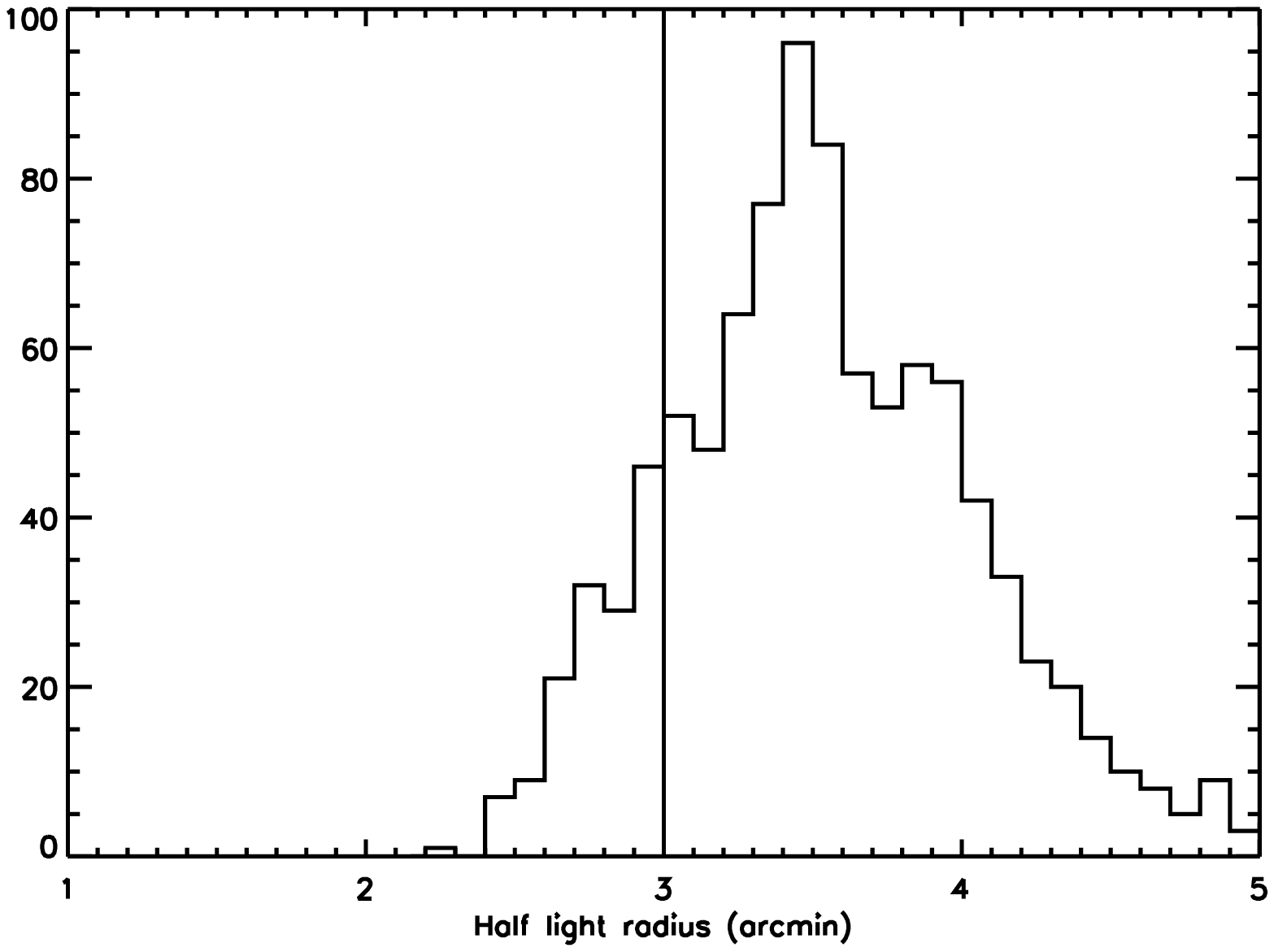}
\epsfysize=4.0cm \epsfbox{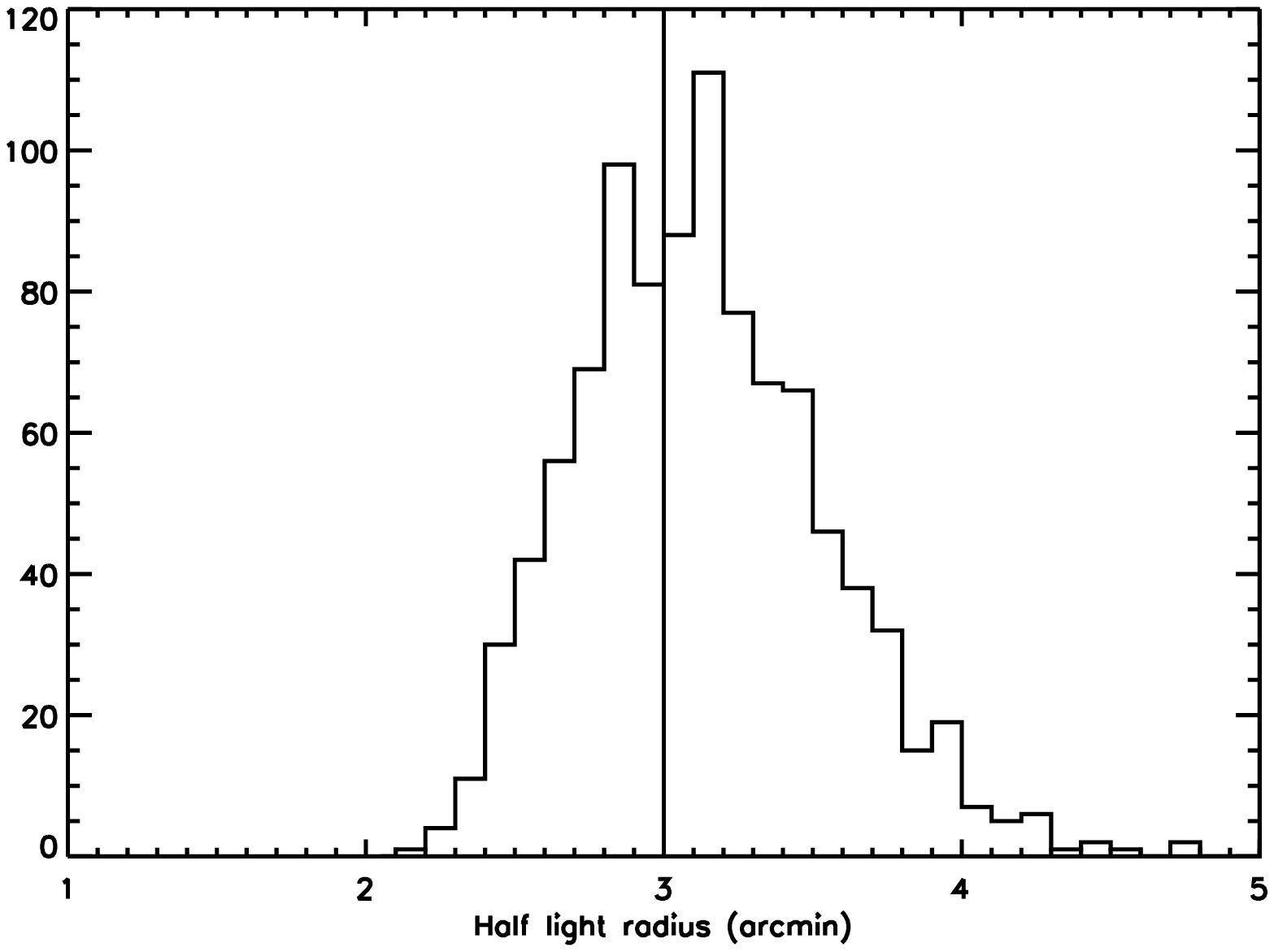}
\epsfysize=4.0cm \epsfbox{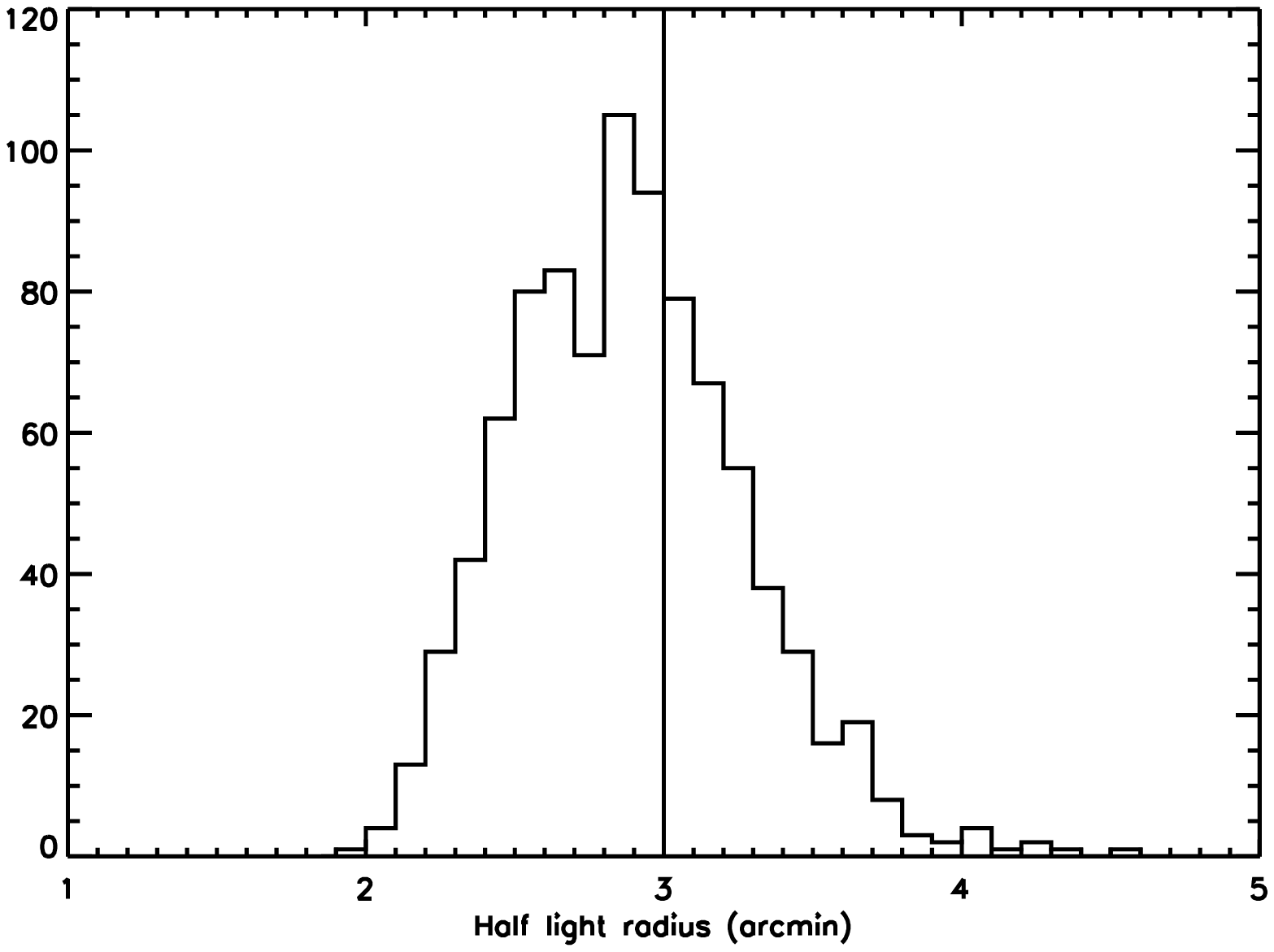}
}
\caption{ An illustration of how the maximum likelihood code for
measuring parameterized structure performs for an input position angle
of 35 degrees while varying the input ellipticity between 0.1 and
0.35.  The output histograms are the results of the one thousand
bootstrap resamples for the labeled scenarios, while the solid
vertical line is the input value of the model, in this case PA = 35.0
degrees (top row) and $r_{h}=3.0$ arcmin (bottom row).  Top row --
Recovered position angle for various mock Leo~IV models with
$\epsilon$ between 0.1 and 0.35.  Note how the example with an input
ellipticity of $\epsilon$=0.10 does not have a clearly measured
position angle, and the peak of the distribution does not agree with
the model's position angle.  It is not until an ellipticity of
$\epsilon \sim$0.35 where the model's position angle is well measured
and accurate.  Bottom -- Recovered half light radius while varying the
input ellipticity.  Although the low ellipticity case of $\epsilon =
0.1$ seems to be slightly offset from the input $r_{h}=3.0$ arcmin
model, the recovered half light radius is still within 1-$\sigma$ of
the input value.  In all other cases, the half light radius is
measured with high precision.
\label{fig:pamodelexample}}
\end{center}
\end{figure*}

We also do a good job of measuring the position angle (as long as the
ellipticity is high enough) and the half light radius of our mock
Leo~IV-like systems, which we illustrate in a series of examples in
Figure~\ref{fig:pamodelexample}.  From the figure, one can see the
gradual improvement in the measurement of the position angle as one
goes from an ellipticity of 0.1 to 0.35, while the half light radius
remains relatively well measured throughout.  The slight systematic
offset seen in the bottom left panel of
Figure~\ref{fig:pamodelexample} can be explained due to the difficulty
in recovering the true ellipticity for $\epsilon_{input} \lesssim
0.25$ systems. If one slightly overestimates the true input
ellipticity of the data, this leads to a slight overestimation of the
half-light radius, a degeneracy that can be seen in Figure 9 of
\citet{sdssstruct}.  The take away message is that one must treat any
'measurement' of the position angle with extreme caution for
ellipticity values of $\epsilon \lesssim 0.25$ as we do in
\S~\ref{sec:structparams}.

In the future, we intend to present a more extensive series of tests
of our maximum likelihood code in order to understand how star number
and imaging field of view have on the estimation of structural
parameters for the new MW satellites.

\clearpage

\bibliographystyle{apj}
\bibliography{apj-jour,mybib}

\clearpage

\clearpage

\clearpage

\begin{inlinefigure}
\begin{center}
\resizebox{\textwidth}{!}{\includegraphics{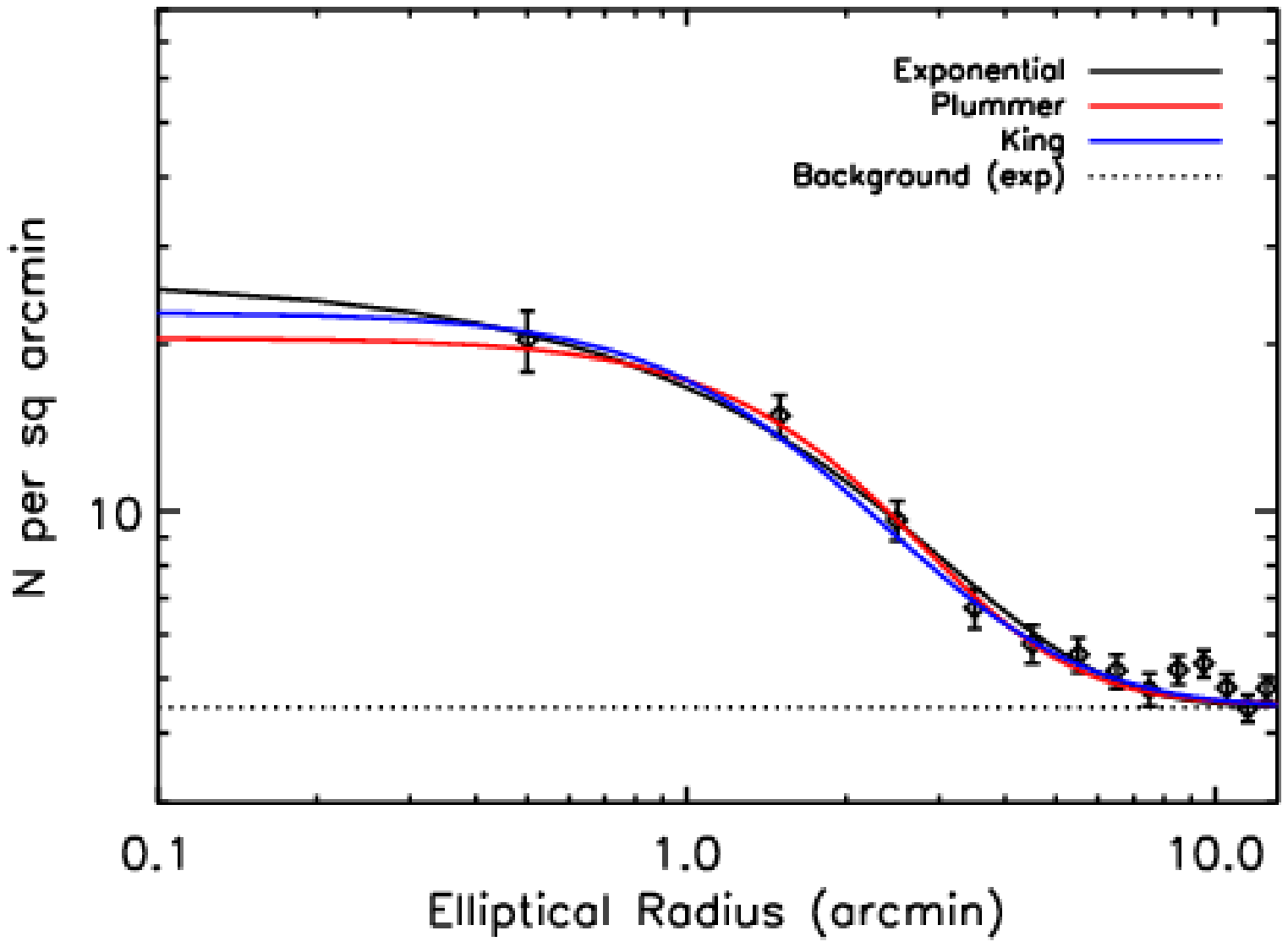}}
\end{center}
\figcaption{Stellar profile of Leo IV.  The data points are the binned
star counts for all stars within the CMD selection box shown in
Figure~\ref{fig:CMDselect}.  The plotted lines show the best fit
one-dimensional exponential, Plummer and King profiles.  The dotted
line shows the background surface density determined for our
exponential profile fit.  Note that in deriving these best fits, we
are not fitting to the binned data, but directly to the stellar
distribution.
\label{fig:stellarprofile}}
\end{inlinefigure}

\clearpage

\clearpage

\clearpage

\clearpage

\clearpage

\clearpage

\begin{inlinefigure}
\begin{center}
\resizebox{\textwidth}{!}{\includegraphics{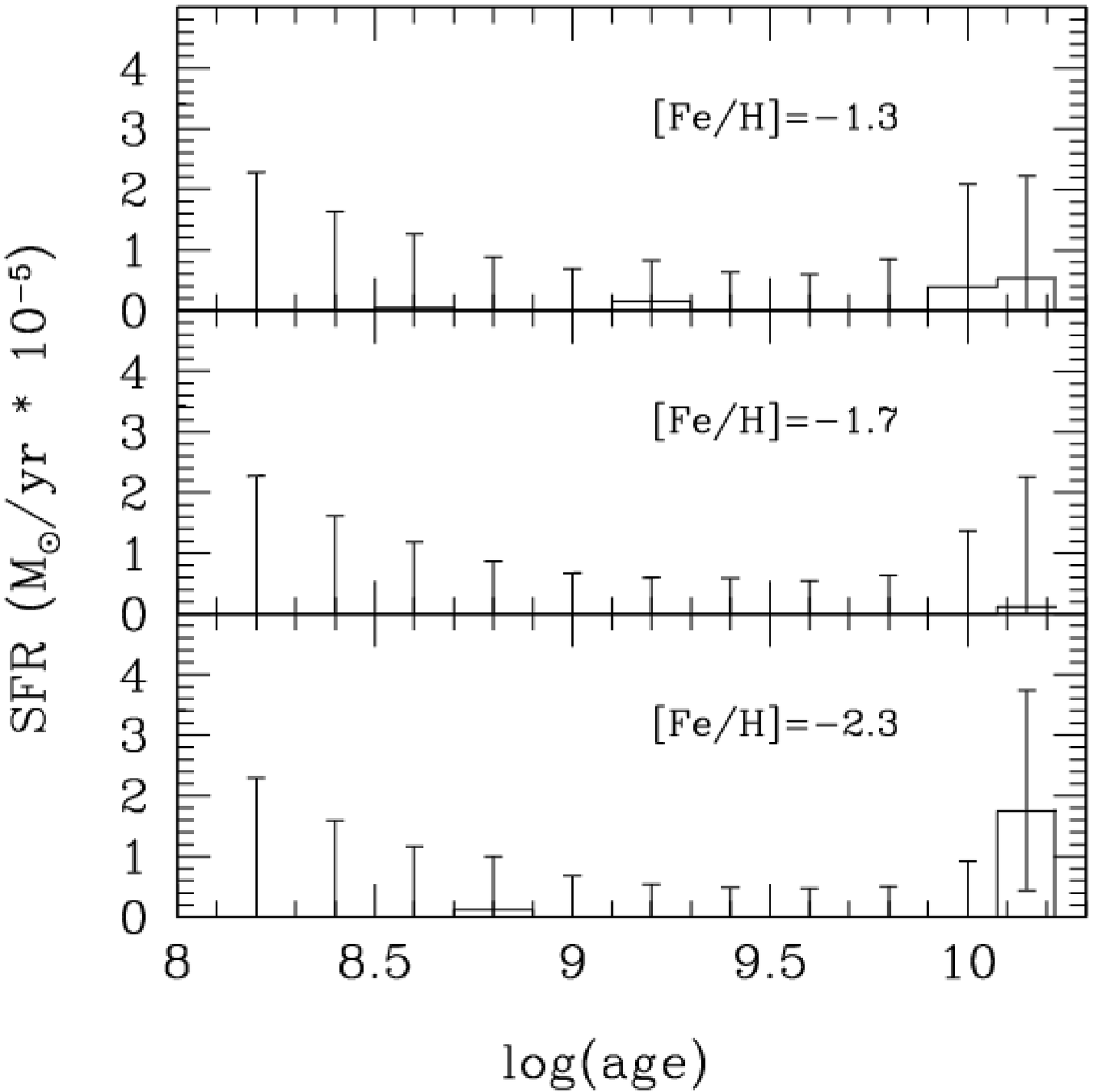}}
\end{center}
\figcaption{SFH solution from the StarFISH fit. With the data in hand,
Leo~IV is consistent with a single, old ($\sim$14 Gyr) stellar
population with [Fe/H]$\sim$-2.3 -- although there is some weak
evidence for a spread in metallicity in this old stellar component.
We also can not rule out a small, young population of stars with our
StarFISH analysis.  Error bars with no accompanying histogram are
upper limits.\label{fig:sfh}}
\end{inlinefigure}

\clearpage

\clearpage

\begin{inlinefigure}
\begin{center}
\resizebox{\textwidth}{!}{\includegraphics{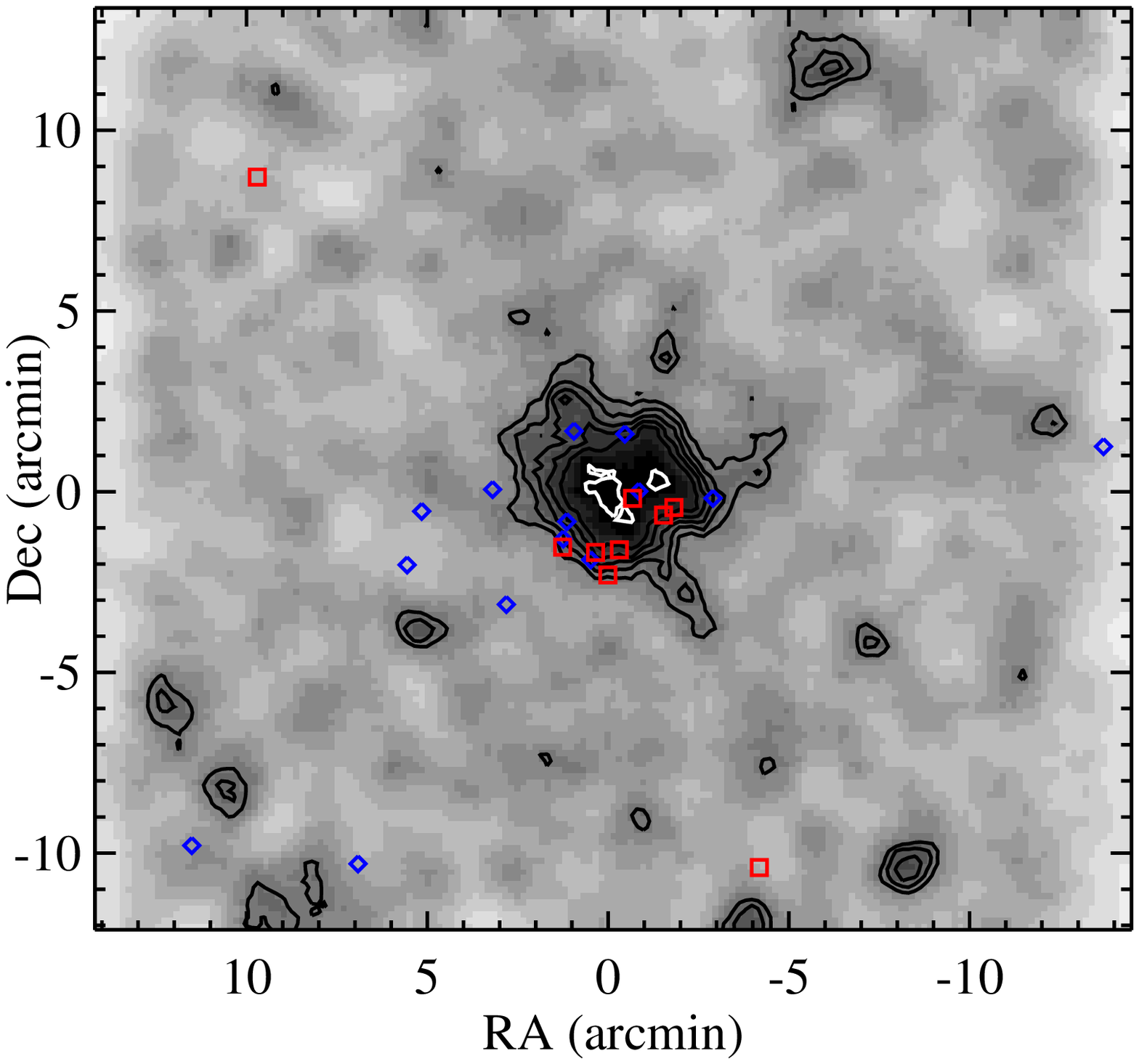}}
\end{center}
\figcaption{ The position of our high-probability blue plume (squares)
and candidate BHB (diamonds) stars overplotted onto the smoothed map
of Leo~IV.  Note that the probable blue plume stars are segregated
within the body of Leo~IV, as would be expected if they represented a
young stellar population rather than blue straggler stars.  As
expected, there are few of these objects outside the main body of
Leo~IV.  The BHB star population is more uniformly distributed across
Leo~IV.
\label{fig:showstars}}
\end{inlinefigure}

\clearpage

\begin{inlinefigure}
\begin{center}
\resizebox{\textwidth}{!}{\includegraphics{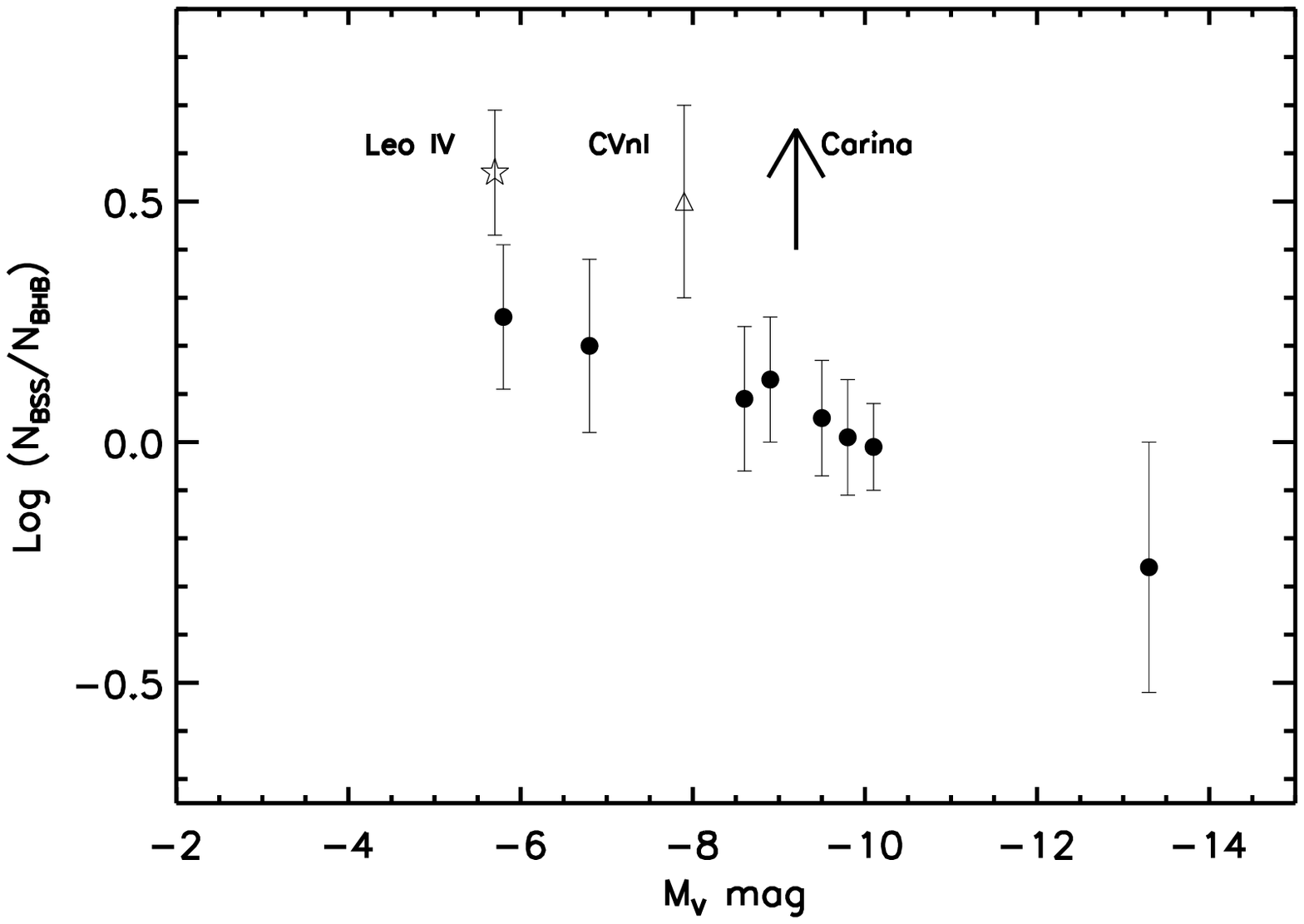}}
\end{center}
\figcaption{ The frequency of blue plume stars, normalized by BHB
stars, for the MW dwarf spheroidals.  The solid, circular points are
the blue plume frequency points for the MW dwarfs derived by
\citet{Momany07}.  Note the anti-correlation between satellite
brightness versus blue plume frequency.  An outlier from that work is
Carina, which likely harbors a young stellar population.  The Canes
Venatici I point is from \citet{Martin08} who reported a young stellar
population in that system due to both the high blue plume frequency
and the segregation of this population within the dwarf.  Likewise,
the blue plume frequency of Leo~IV lies off the relation of
\citet{Momany07}.  That, along with the segregation seen in
Figure~\ref{fig:showstars} indicates that Leo~IV harbors a small,
young stellar population.
\label{fig:bpfreq}}
\end{inlinefigure}

\clearpage

\begin{inlinefigure}
\begin{center}
\resizebox{\textwidth}{!}{\includegraphics{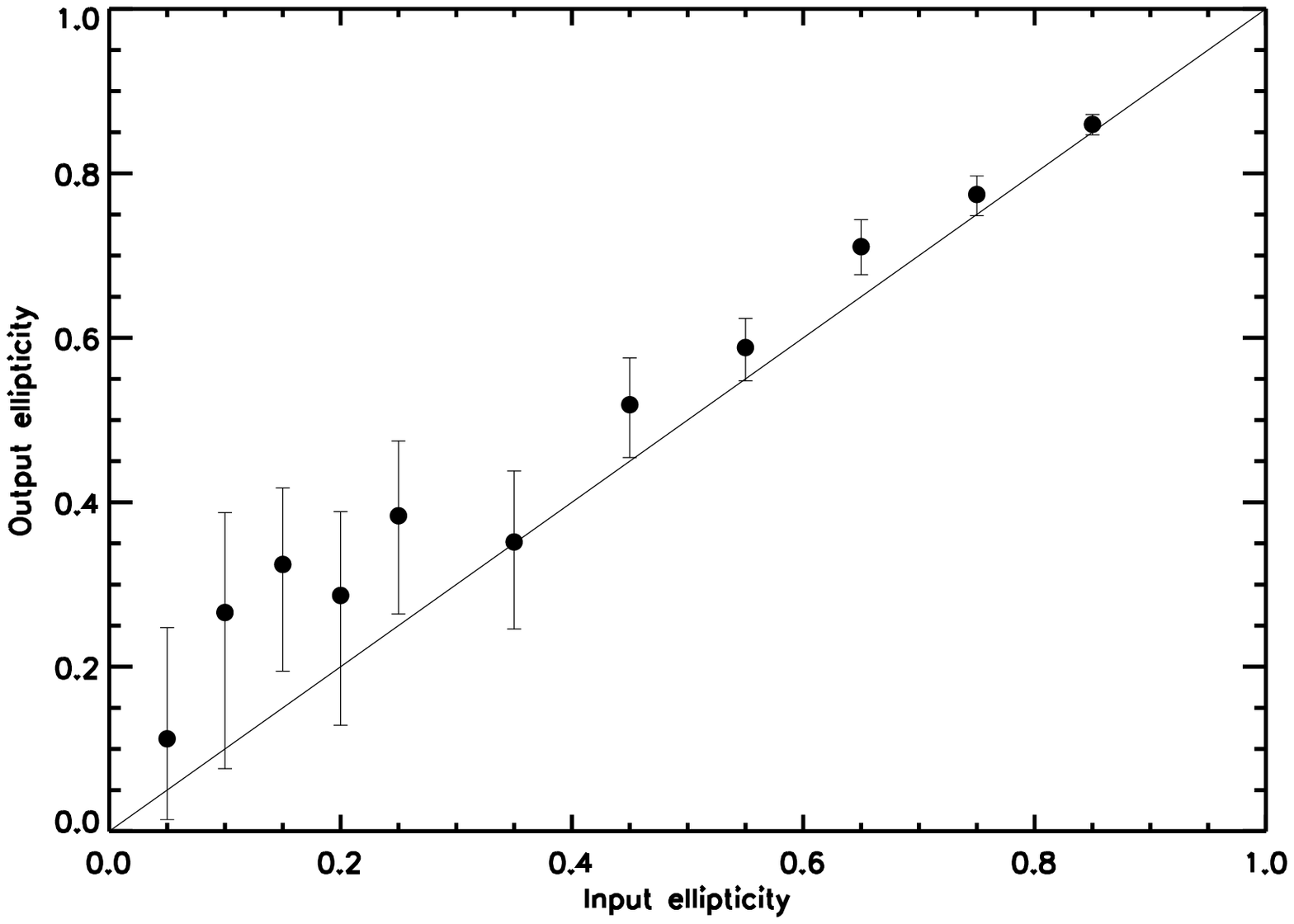}}
\end{center}
\figcaption{ Comparison of the input ellipcity in our mock Leo~IV
dwarf galaxy models versus the output ellipticity found by our maximum
likelihood algorithm.  The points represent the median value of the
1000 bootstrap resamples, and the error bars represent the central
68.3\% of the distribution around the median.  The dashed line is the
one-to-one correspondence between input and output.  For input
ellipticity values of $\epsilon_{input} \lesssim 0.25$ we
systematically overpredict the ellipticity with large error bars and
thus only quote upper limits.
\label{fig:ellinout}}
\end{inlinefigure}

\clearpage

\clearpage

\begin{deluxetable}{lcccccccccc}
\tablecolumns{11}
\tablecaption{Leo IV Photometry \label{table:phot}}
\tablehead{
\colhead{Star No.} & \colhead{$\alpha$} & \colhead{$\delta$} & \colhead{$g$} &\colhead{$\delta g$} & \colhead{$A_{g}$} & \colhead{$r$} &\colhead{$\delta r$} & \colhead{$A_{r}$} &\colhead{SDSS or MMT}\\
\colhead{} & \colhead{(deg J2000.0)} & \colhead{(deg J2000.0)} & \colhead{(mag)}&\colhead{(mag)} & \colhead{(mag)} & \colhead{(mag)} & \colhead{(mag)} & \colhead{(mag)} &\colhead{}}
\startdata
0&173.23115&-0.54635&18.23&0.02&0.09&17.93&0.01&0.07&SDSS\\
1&173.24960&-0.50828&17.16&0.01&0.09&16.61&0.01&0.06&SDSS\\
2&173.20208&-0.54312&17.03&0.02&0.09&16.61&0.01&0.07&SDSS\\
3&173.22927&-0.49215&18.66&0.02&0.09&17.20&0.01&0.06&SDSS\\
4&173.25937&-0.50253&17.87&0.01&0.09&17.47&0.01&0.06&SDSS\\
5&173.19711&-0.49921&16.58&0.02&0.09&15.84&0.01&0.06&SDSS\\
6&173.21481&-0.58547&16.52&0.02&0.09&15.50&0.01&0.07&SDSS\\
7&173.23169&-0.47729&17.00&0.02&0.09&15.81&0.01&0.06&SDSS\\
8&173.23995&-0.60117&18.08&0.02&0.09&17.47&0.01&0.07&SDSS\\
9&173.19902&-0.56584&19.31&0.02&0.09&17.78&0.01&0.07&SDSS\\
\enddata
\tablenotetext{a}{See electronic edition for complete data table.}
\end{deluxetable}

\clearpage

\begin{deluxetable}{lccc}
\tabletypesize{\small}
\tablecolumns{10}
\tablecaption{Leo IV structure -- parameterized fits \label{table:paramfits}}

\tablehead{
\colhead{Parameter} & \colhead{Measured} & \colhead{Uncertainty} & \colhead{Bootstrap median}\\
}
\startdata

$M_{V}$ & -5.5 & 0.3 & -5.5 \\
$\mu_{0,V}$ & 27.2 & 0.3 & 27.2 \\ 
\hline
Exponential Profile \\
\hline\hline
RA (h~m~s)&11:32:56.38&$\pm14''$&11:32:56.33\\
DEC (d~m~s)&-00:32:27.25&$\pm14''$&-00:32:29.24\\
$r_{h}$ (arcmin) &2.85&0.64&3.17\\
(pc) & 127.8 & 28.8 & 142.2 \\
$\epsilon$&0.05&$<0.23$&0.14\\
\hline
Plummer Profile \\
\hline\hline
RA (h~m~s)&11:32:56.20&$\pm12''$&11:32:56.18\\
DEC (d~m~s)&-00:32:27.17&$\pm10''$&-00:32:28.32\\
$r_{h}$ (arcmin) &2.86&0.40&2.97\\
(pc) &128.1&18.0&133.3\\
$\epsilon$&0.02&$<0.20$&0.11\\
\hline
King Profile \\
\hline\hline
RA (h~m~s)&11:32:56.04&$\pm13''$&11:32:56.08\\
DEC (d~m~s)&-00:32:29.60&$\pm10''$&-00:32:29.95\\
$r_c$ (arcmin)&1.61&0.22&1.64\\
(pc)&72.2&10.3& 73.4\\
$r_t$ (arcmin)&18.55&4.39&20.25\\
(pc)&831.0&196.7&907.3\\
$\epsilon$&0.03&$<0.18$&0.09\\
\hline\hline
\\

\enddata \tablenotetext{a}{All transverse distances are reported using
a $(m-M)$ = 20.94 distance modulus.}
\tablenotetext{b}{Absolute magnitude and central surface brightness are calculated using the exponential profile fit.}
\tablenotetext{c}{The value in the uncertainty column for $\epsilon$
  corresponds to the 68\% upper confidence limit, given that our
  derived $\epsilon$ is consistent with 0.}
\end{deluxetable}

\clearpage

\begin{deluxetable}{lccccc}
\tabletypesize{\small}
\tablecolumns{10}
\tablecaption{Simulated Leo~IV External Structures and Detections \label{tab:fakeresult}}

\tablehead{
\colhead{No. of stars} & \colhead{$M_{r}$} & \colhead{$\mu_{0,r}$} & \colhead{$M_{g}$} & \colhead{$\mu_{0,g}$} & \colhead{Peak $\sigma$} \\
}
\startdata
Input 'Nuggets' \\
\hline
\hline
25 & -1.0 & 29.7 & -0.6 & 30.1 & 4.0 \\
35 & -2.8 & 27.9 & -2.5 & 28.3 & 6.4 \\
50 & -3.2 & 27.5 & -3.0 & 27.7 & 9.7 \\
\hline
Input 'Streams' \\
100 & -3.5 & 30.5 & -3.2 & 30.8 & 1.7 \\
200 & -4.4 & 29.6 & -4.3 & 29.8 & 4.3 \\
300 & -5.4 & 28.7 & -5.0 & 29.0 & 5.1 \\
\hline
\hline

\enddata \tablenotetext{a}{All nuggets have an exponential profile
with half light radius of 1 arcminutes.}  \tablenotetext{b}{All
streams have a constant density in the declination direction, with a
gaussian density profile in the right ascension direction, with
$\sigma$=1.5 arcmin. } \tablenotetext{c}{We quote the peak $\sigma$
for our artificial streams at a position of $(+0.0,+8.0)$ arcminutes
with respect to the center of Leo~IV}
\end{deluxetable}

\end{document}